# A Survey on Security and Privacy Protocols for Cognitive Wireless Sensor Networks

Jaydip Sen

Department of Computer Science and Engineering
National Institute of Science and Technology, Odisha, INDIA
Jaydip.Sen@nist.edu

**Abstract:** Wireless sensor networks (WSNs) have emerged as an important and new area in wireless and mobile computing research because of their numerous potential applications that range from indoor deployment scenarios in home and office to outdoor deployment in adversary's territory in a tactical battleground. Since in many WSN applications, lives and livelihoods may depend on the timeliness and correctness of sensor data obtained from dispersed sensor nodes, these networks must be secured to prevent any possible attacks that may be launched on them. Security is, therefore, an important issue in WSNs. However, this issue becomes even more critical in *cognitive wireless sensor networks* (CWSNs), a type of WSNs in which the sensor nodes have the capabilities of changing their transmission and reception parameters according to the radio environment under which they operate in order to achieve reliable and efficient communication and optimum utilization of the network resources. This survey paper presents a comprehensive discussion on various security issues in CWSNs by identifying numerous security threats in these networks and defense mechanisms to counter these vulnerabilities. Various types of attacks on CWSNs are categorized under different classes based on their natures and targets, and corresponding to each attack class, appropriate security mechanisms are presented. The paper also identifies some open problems in this emerging area of wireless networking research.

## 1. Introduction

Over the last decade, wireless sensor networks (WSNs) have attracted a lot of interest in the research community due to their wide range of potential applications. A WSN consists of hundreds or even thousands of small devices each with sensing, processing, and communication capabilities to monitor a real-world environment. They are envisioned to play an important role in a wide variety of areas ranging from critical military surveillance applications to forest fire monitoring and building security monitoring (Akyildiz et al., 2002). Most of the WSN deployments operate in the unlicensed ISM bands (2.4GHz). Several other small range wireless protocols like Wi-Fi, Bluetooth etc. also use the same band. This has led to overcrowding in this band with the increasing deployment of WSN-based applications. As a result, coexistence issues in the ISM bands have attracted extensive research attention (Howitt & Gutierrez, 2003; Cavalcanti et al., 2007).

The increasing demand for spectrum in wireless communication has made efficient spectrum utilization a big challenge. To address this important requirement, *cognitive radio* (CR) has emerged as the key technology. A CR is an intelligent wireless communication system that is aware of its surrounding environment, and adapts its internal parameters to achieve reliable and efficient communication and optimum utilization of the resources (Mitola, 2000).

With the advent of CR technology, we have a different perspective of the traditional WSNs. In the current cognitive wireless sensor networks (CWSNs), the nodes change their transmission and reception parameters according to the radio environment. Cognitive capabilities are based on four activities: (i) monitoring of spectrum sensing, (ii) analysis and characterization of the environment, (iii) optimization of the best communication strategy based on different constraints such as reliability, power, security and privacy issues etc., and (iv) adaptation and collaboration strategy. The cognitive technology will not only enable access to new spectrum but it will also provide better propagation characteristics leading to reduction in power consumption, network life-time and reliability in a WSN. With cognitive capabilities, WSNs will be capable of finding a free channel in the unlicensed band to transmit or could find a free channel in the licensed band for communication. A CWSN, therefore, will be able to provide access not only to new spectrum bands in addition to the available 2.4 GHz band, but also to the spectrum band that has better propagation characteristics. If a channel in a lower frequency band is accessed, it will certainly allow communications with higher transmission range in a CWSN, and hence fewer sensor nodes will be required to provide coverage in a specific area with a higher network life-time due to lower energy consumption in the nodes. CWNs will also provide better propagation characteristics by adaptively changing system parameters like modulation schemes, transmit power, carrier frequency and constellation size. The result will be a more reliable communication with reduced power consumption, increased network life-time and higher reliability and enhanced *quality of service* (QoS) guarantee to applications.

Although there are several advantages and benefits that can be achieved by deploying CWSNs (Cavalcanti et al., 2008), guaranteeing security poses a significant challenge. Unless these challenges are solved to an effective level, deployment of CWSNs in real-world applications may face a serious impediment. As observed in (Burbank, 2008), the CR nature of a system introduces an entirely new gamut of threats and vulnerabilities that cannot be easily mitigated. The three salient characteristics of CR are its environmental awareness, learning and acting capabilities. Considering these characteristics from an attacker's perspective, a CWSN will provide much more capability to an attacker to launch attacks that are long-lasting and catastrophic in nature and those which can be triggered by simple spectral manipulations (Araujo, et al., 2012).

Security had already been an extensive area of research in WSNs (Sen, 2009; Du & Chen, 2008; Walters et al., 2006; Yong et al., 2006; Zhou et al., 2008; Martins & Guyennet, 2010). With the advent of CWSNs and the perspective of security taking a much wider and complicated scope, it is obvious that research on the security aspects on CWSNs will attract even more attention of the research community. At present, however, despite considerable amount of ongoing research on CR networks (Clancy & Goergen, 2008), and the new interest in CWSNs (Zahmati et al., 2009) security in CWSNs has been a vastly unexplored area. Preservation of sensor data privacy is also a critical issue when these networks are deployed for applications that deal with sensitive and critical data.

This paper provides a panoramic view of the security and privacy-related issues in WSNs with a particular focus on CWSNs. The rest of the paper is organized as follows. Section 2 presents a brief discussion on various security and privacy issues in a traditional WS, which are applicable to CWSNs as well. Section 3 discusses some of the security and defense mechanisms for tackling these vulnerabilities. In Section 4, we discuss security vulnerabilities which are specific to CWSNs. Section 5 presents some of attacks on CWSNs based on the current state of the art. In Section 6, we discuss various security mechanisms for defending against attacks on CWSNs. Section 7 identifies some future research challenges on security and privacy issues in CWSNs. Finally, Section 8 concludes the paper.

## 2. Security and Privacy Issues in WSNs

Traditional WSNs are vulnerable to various types of attacks. These attacks can be broadly categorized into the following types (Shi & Perrig, 2004):

*Attacks on secrecy and authentication*: standard cryptographic mechanisms can prevent attacks on the secrecy and authenticity of the messages from outsider attacks such as eavesdropping, packet replay attacks, and modification or spoofing of packets.

*Attacks on network availability*: these attacks are more generally known as the *denial of service* (DoS) attacks and they can be launched on any layer of the communication protocol stack.

*Stealthy attacks against service integrity:* in these attacks, the goal of the attackers is to lure the network accept a false data value. For example, an attacker compromises a sensor node and injects a false data value through that sensor node.

In the following, we discuss various types of attacks in detail. First, we describe various ways in which the DoS attacks can be launched on a traditional WSN. In section 4 we present various possible attacks on CWSNs. It must be understood, however, that all vulnerabilities of traditional WSNs are applicable to CWSNs as well.

### 2.1 Denial of Service (DoS) Attacks

Wood et al. define a DoS attack as an event that diminishes or attempts to reduce the capacity of network to perform its desired function (Wood & Stankovic, 2002). In the following, we describe how DoS attacks can be launched in the different layers of the communication protocol stack in a traditional WSNs.

### 2.1.1 DoS attacks on the physical layer

The physical layer is responsible for frequency selection, carrier frequency generation, signal detection, modulation, and data encryption (Akyildiz et al., 2002). Jamming in the physical layer is the most usual way to launch a DoS attack. In the jamming attack, the attacker interferes with the radio frequencies that the nodes in a WSN use for communication (Wood & Stankovic, 2002; Shi & Perrig, 2004). The jamming attack is extremely catastrophic. Even with a less powerful jamming source, an adversary can potentially disrupt communication in an entire network by strategically distributing the sources of the jamming signal.

### 2.1.2 DoS attacks on the link layer

The link layer is responsible for multiplexing of data streams, data frame detection, medium access control, and error control (Akyildiz et al., 2002). The attacks launched on this layer usually create collisions, resource exhaustion, and unfairness in allocation. A collision occurs when two nodes attempt to transmit simultaneously on the same frequency. An adversary may strategically cause collisions in specific packets such as ACK control messages. A possible result of such collisions is the costly exponential back-off. Repeated collisions of frames may lead to resource exhaustion in the sensor nodes (Wood & Stankovic, 2002).

### 2.1.3 DoS attacks on the network layer

The network layer of traditional WSNs is vulnerable to different types of attacks such as *spoofed routing information*, *selective packet forwarding*, *sinkhole*, *Sybil*, *wormhole*, *blackhole*, *grayhole*, *HELLO flood*, *Byzantine*, *information disclosure*, and *acknowledgement spoofing*. In the spoofed routing information attack, an attacker targets the routing information in the network by spoofing, altering or replaying the routing information to disrupt the traffic in the network (Hoffstein et al., 1998). The disruptions include creation of routing loops, attracting or repelling the network traffic from selected nodes, extending or shortening the source routes, generating fake error messages, causing network partitioning, and increasing the end-to-end latency. In the selective forwarding attack, the attacker compromises a node in such a way that it selectively forwards some messages and drops the others (Wang et al., 2009a). In a sinkhole attack, an attacker makes a compromised node more attractive to its neighbors by forging routing information (Wood & Stankovic, 2002; Karlof & Wagner, 2003; Newsome et al., 2004). The result is that the neighbor nodes choose the compromised nodes as the next-hop node to route their data through. This type of attack makes selective forwarding very simple as all traffic from a large area in the network flows through the compromised node. In the Sybil attack, a malicious node presents more than one identity in a network. This attack is particularly effective on routing algorithms, data aggregation, voting, fair resource allocation, and misbehavior detection. For instance, in a sensor network voting scheme, a Sybil attack might utilize multiple identities to generate additional "votes". Similarly, to attack a routing protocol, the Sybil attack can rely on a malicious node taking on the identities of multiple nodes and routing packets through a single malicious node. In the wormhole attack, a pair of malicious nodes first creates a wormhole. A wormhole is a low-latency link between two portions of a network over which one attacker node replays messages to the other attacker node (Karlof & Wagner, 2003). This link may be established either by a single node forwarding messages between two adjacent but otherwise non-neighboring nodes or by a pair of nodes in different parts of a network communicating with each other. The latter case is closely related to the sinkhole attack as an attacking node near the base station can provide a one-hop link to that base station via the other attacking node in a distant part of the network. In the blackhole attack, a malicious node falsely advertises good paths (e.g., the shortest path or the most stable path) to the destination node during the path-finding process in reactive routing protocols, or in the route update messages in proactive routing protocols. The intention of the malicious node could be to hinder the path-finding process or to intercept all data packets sent to the concerned destination node. A more delicate form of this attack is known as the grayhole attack, in which the malicious node intermittently drops data packets thereby making its detection more difficult. In an HELLO flood attack, an attacker may use a high-powered transmitter to fool a large number of nodes and make them believe that they are within its neighborhood (Karlof & Wagner, 2003). Subsequently, the attacker node falsely broadcasts a shorter route to the base station and all the nodes that received the HELLO packets attempt to transmit to the attacker node. However, since these nodes are out of the radio range of the attacker, no communication will be established. In the Byzantine attack, a single compromised node or a set of compromised nodes works in collusion and carries out attacks by creating routing loops, forwarding packets through suboptimal routes, and selectively dropping packets (Awerbuch et al., 2002). These attacks are very difficult to detect since under such attacks, the networks usually do not exhibit any abnormal behavior. In an information disclosure attack, a compromised node leaks confidential or important information to unauthorized nodes in the network. Such information may include information regarding the network topology, geographic location of nodes, or optimal routes to authorized nodes in the network. In resource depletion attack, a malicious node attempts to deplete resources of other nodes in the network. The typical resources that are targeted are battery power, bandwidth, and computational power. The attacks could also be in the form of unnecessary requests for routes, very frequent generation of beacon packets, or forwarding of stale packets to other nodes. The acknowledgment spoofing attack is launched on routing algorithms that require transmission of acknowledgment packets. An attacking node may overhear packet transmissions from its neighboring nodes and spoof the acknowledgments, thereby providing false information to the nodes (Karlof & Wagner, 2003). In this way, the attacker is able to disseminate wrong information in the network about the status of the nodes, since acknowledgments may arrive from nodes that are not alive in reality.

In addition to the aforementioned categories of attacks, various other types of attacks are possible on the routing protocols in WSNs. Most of the routing protocols in WSNs are vulnerable to attacks such as routing table overflows, routing table poisoning, packet replication, route cache poisoning, and rushing attacks. A comprehensive discussion on these attacks may be found in (Sen, 2010a).

**Table 1**. Various types of DoS attacks and their possible countermeasures in WSNs

| Layer | Attacks | Defense Mechanisms |
|---|---|---|
| Physical | Jamming | Spread-spectrum, priority messages, lower duty cycle, region mapping, mode change |
| | Collision | Error-correction code |
| MAC | Exhaustion | Rate limitation |
| | Unfairness | Small frames |
| Network | Spoofed routing information & selective forwarding | Egress filtering, authentication, monitoring |
| | Sinkhole | Redundancy checking |
| | Sybil | Authentication, monitoring, redundancy |
| | Wormhole | Authentication, probing |
| | Hello Flood | Authentication, packet leashes by using geographic and temporal information |
| | Ack. Flooding | Authentication, bi-directional link authentication verification |
| Transport | SYN Flooding | Client puzzles, SSL-TLS |
| | De-synchronization | authentication, EAP |
| Application | Logic errors | Application authentication |
| | Buffer overflow | Trusted computing, Antivirus |
| Privacy | Traffic analysis, Attack on data privacy and location privacy | Homomorphic encryption, Onion routing, schemes based on traffic entropy computation, group signature based anonymity schemes, use of pseudonyms. |

### 2.1.4 DoS attacks on the transport layer

The attacks that can be launched on the transport layer of a WSN communication protocol stack are the *flooding attack* and the *desynchornization attack*. If a protocol needs to maintain the state information at either end of an established connection, it becomes vulnerable to memory exhaustion attack (Wood & Stankovic, 2002). An attacker may repeatedly make new connection requests until the resources required by each connection are exhausted or reach a maximum limit. In either case, further legitimate requests are ignored by the victim node.

The desynchornization attack, on the other hand, attempts to disrupt an existing connection (Wood & Stankovic, 2002). An attacker may, for example, repeatedly spoof messages to an end host causing the host to request retransmission of missed frames. If timed correctly, an attacker may degrade or even prevent the ability of end hosts to successfully exchange data, causing them to waste energy instead of attempting to recover from errors that never really exist. The possible DoS attacks on WSNs and their corresponding countermeasures are listed in Table 1.

### 2.1.5 Attacks on secrecy and authentication

There are different types of attacks under this category. We mention only the node replication attack. A more detailed discussion can be found in (Sen, 2009).

*Node replication attack:* In this attack, the attacker attempts to add a node to an existing WSN by replicating (i.e. illegally copying) the node identifier of an already existing node in the network (Parno et al., 2005). A node replicated and joined in the network in this manner can potentially cause severe disruption in message communication in the WSN by corrupting the packets and forwarding them to wrong routes. This may also lead to network partitioning and communication of false sensor readings. In addition, if the attacker gains a physical access to the network, it is possible for him/her to copy the cryptographic keys and use these keys for message communication from the replicated node. The attacker may also place the replicated node in strategic locations in the network so that he/she could easily manipulate a specific segment of the network, possibly causing a network partitioning.

### 2.1.6 Attacks on sensor data privacy

Since in many applications WSNs are deployed for automatic data collection through efficient and strategic deployment of the sensor nodes, these networks are vulnerable to potential abuse of the collected data. Privacy preservation of sensitive data in WSNs is a particularly difficult challenge (Gruteser et al., 2003). Moreover, an adversary may gather seemingly innocuous data to derive sensitive information if he or she knows how the aggregate data is collected from multiple sensor nodes. This is analogous to the "*panda hunter problem*", in which the hunter can accurately estimate the location of the panda by systematically monitoring the traffic (Ozturk et al., 2004). Some of the common attacks on sensor data privacy (Gruteser et al., 2003; Chan & Perrig, 2003) are briefly discussed in the following.

- *Eavesdropping and passive monitoring*: The most common form of attack on sensor data privacy is carried out by an attacker by silently listening to the messages communicated over the network. If the messages are not protected using cryptographic mechanisms, the adversary can easily understand their contents.

- *Traffic analysis*: In order to launch an attack on privacy, an attacker sometimes combines passive eavesdropping with an active traffic analysis. Through an effective analysis of traffic, an adversary can identify some sensor nodes with special roles and activities in a WSN.

- *Camouflage*: In camouflage attack, an adversary compromises a sensor node and later on uses the victim node to masquerade as a normal node in the network. This camouflaged node may advertise false routing information and attract packets from other nodes for further forwarding. After the packets start arriving at the compromised node, it starts forwarding them to strategic nodes where privacy analysis of the packets may be carried out systematically.

## 3. Security Mechanisms in Traditional WSNs

Numerous security mechanisms have been proposed by the researchers for defending against the possible attacks on WSNs. In the following, we provide a very brief discussion on some of the well known defense mechanisms for WSNs without aiming to present a comprehensive discussion on any of these schemes. Interested readers may refer to (Sen, 2009) for a detailed discussion.

### 3.1 Applications of cryptographic mechanisms

Since most of the security mechanisms for WSNs use cryptography, selecting the most appropriate cryptographic mechanism is a critical issue. The cryptographic algorithms and protocols must meet the constraints of the sensor nodes and should be evaluated by their code sizes, data sizes, processing time, and computational power requirements. It was popular belief for long that the code size, processing time, and power requirements of the public key algorithms such as Diffie-Hellman key exchange protocol (Malan et al., 2004) or RSA signatures (Rivest et al., 1978) are too high for WSN nodes. However, subsequent studies have shown that it is feasible to apply public key cryptography in WSNs by right selection of algorithms and associated parameters, optimization, and the use of low-power techniques (Gura et al., 2004; Gaubatz et al., 2004; Wander et al., 2005). For example, the public key algorithms like Rabin's scheme (Rabin, 1979), Ntru-Encrypt (Hoffstein et al., 1998), RSA (Rivest et al., 1978), and the *elliptic curve cryptography* (ECC) (Miller, 1986; Kobiltz, 1987) are all found to be feasible in WSN applications. ECC is particularly suitable for WSNs since it provides the same level of security as the RSA algorithm with a far smaller key size, thereby reducing the processing and communication overhead. In general, however, the private key operations in the public key cryptographic schemes are still expensive and most of the private key-related operations are assumed to be either carried out by the base stations or on some selected sensor nodes which have higher computational resources (Malan et al., 2004; Rivest et al., 1978; Brown et al., 2000; Gura et al., 2004; Gaubatz et al., 2004). Symmetric key-based protocols such as RC4 (Menezes et al., 1996), RC5 (Rivest, 1995), IDEA (Menezes et al., 1996), SHA-1 (Eastlake & Jones, 2001) and MD5 (Menezes et al., 1996; Rivest, 1992) are also widely used for ensuring message authentication, confidentiality, and integrity in WSNs.

### 3.2 Key management protocols

Since the existence of a robust and efficient key management protocol is an essential pre-requirement for successful operation of a cryptographic mechanism, design of attack-resilient key management schemes that meet the resource constraints in such networks is a challenging task. The goal of key management is to establish keys among the nodes in a secure and reliable manner and to support node addition and revocation. Due to the high computational overhead of most of the public key cryptosystems, majority of the existing key management schemes for WSNs are based on symmetric key cryptography (Sen, 2009). A large number of key management protocols for WSNs have been proposed by the researchers. A comprehensive discussion on key management in WSNs can be found in (Sen, 2009).

### 3.3 Defense mechanisms against DoS attacks

Since DoS attacks can be launched at different layers of the protocol stack, the defense mechanisms at different layers follow different approaches.

In the physical layer, jamming attack can be defended by employing variations of spread-spectrum communications such as *frequency hopping* and *code spreading* (Wood & Stankovic, 2002). In *frequency-hopping spread spectrum* (FHSS), signals are transmitted by rapidly switching a carrier among many frequency channels using a pseudo-random sequence that is known to both the transmitter and the receiver. As a potential attacker would not be able to predict the frequency selection sequence, it will be impossible for him/her to jam the frequency being used at a given point of time. Another approach for handling jamming attacks in WSN is to tolerate the attacks by correctly identifying the jammed part of the network and effectively avoiding the nodes in the affected part by routing messages around it. Wood et al. (Wood & Stankovic, 2002) have proposed an approach in which nodes along the perimeter of a jammed region report their status to their neighbors and the affected region is identified collectively and packets are routed around it.

In the link layer, *frame collision attacks* are handled by using error-correcting codes (Wood & Stankovic, 2002). The resource (i.e., energy) exhaustion attacks are prevented by *applying rate-limiting admission control mechanism* in the *medium access control* (MAC) layer so that the requests from nodes that intend to exhaust the energy-reserves of a node are rejected. Use of *time-division multiplexing* is another approach to defend against energy exhaustion attacks (Wood & Stankovic, 2002). Time-division multiplexing eliminates the need of arbitration for each frame and solves the indefinite postponement problem in a back-off algorithm. The adverse impact of unfairness caused by an attacker who intermittently launches link layer attacks can be mitigated by the use of small frames since it reduces the amount of time an attacker gets to capture the communication channel (Wood & Stankovic, 2002). However, this technique often reduces the throughput and it is susceptible to further unfairness if that attacker tries to retransmit quickly instead of randomly delaying his/her retransmission attempt.

### 3.4  Defense against attacks on the routing protocols

Numerous mechanisms exist for defending attacks on the network layer and on the routing protocols of WSNs. Since a detailed discussion of these schemes is beyond the scope of this paper, we provide only a very brief discussion on some of the current and popular mechanisms. A detailed discussion can be found in (Sen, 2009).

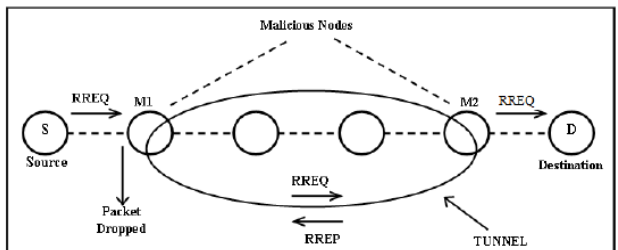

**Figure 1.** An illustration of the *wormhole attack* launched by nodes *M1* and *M2* in a WSN.

A popular way to prevent spoofing and alteration of the routing packets is to append a *message authentication code* (MAC) to the routing packets. To defend against replayed information, counters or time-stamps are used in the messages (Perrig et al., 2002). *Selective forwarding* (or selective packet dropping) attacks may be prevented using *multipath routing* (Karlof & Wagner, 2003). Hu et al. have proposed a mechanism called "*packet leashes*" for detecting and defending against wormhole attacks (Hu et al., 2003). As shown in Figure 1, in a wormhole attack, two or more malicious nodes collude together by establishing a tunnel using an efficient communication medium (i.e., a wired link or a high-speed wireless connection). During the route discovery phase, the route request messages are forwarded between the malicious nodes using the established tunnel. Therefore, the request message that reaches first at the destination node is the one that is forwarded by the malicious nodes. Consequently, the malicious nodes are added in the path from the source to the destination. Once the malicious nodes are included in the routing path, the malicious nodes either drop all the packets, resulting in complete denial of service, or drop the packets selectively to avoid detection.

Sen et al. present a cooperative detection scheme that exploits the redundancy in routing information in an ad hoc network to build a robust detection framework for identifying malicious packet dropping nodes (Sen et al., 2007a). In (Sen et al., 2007b), a cooperative grayhole attack detection mechanism is proposed that utilizes a robust distributed collaborative algorithm among the nodes in an ad hoc network. Di Pietro et al. propose a mechanism for securing group communications in WSNs (Di Pietro et al., 2003). The protocol is known as LKHW (Logical Key Hierarchy for Wireless sensor networks) and it is based on *directed diffusion-based multicast* mechanism. For its operation, the protocol deploys a logical hierarchy that has a central key distributor at the root of the tree and the nodes in the WSN are the leaf level. The internal nodes of the tree contain keys that are used in the rekeying process. Using the directed diffusion approach (Intanagonwiwat et al., 2000), data dissemination in the network is done in an efficient manner. Lazos et al. propose a similar tree-based key distribution scheme in which a routing tree is constructed with the leaf nodes having the keys assigned to them and the nodes at the intermediate levels of the tree acting as the relay nodes (Lazos & Poovendran, 2002).

As discussed earlier in this section, most of the routing protocols for WSNs are vulnerable to various types of attacks such as: selective forwarding, sinkhole, blackhole, grayhole etc. For a detailed discussion on these attacks and a comparative analysis of some of the well-known secure routing protocols for WSNs, the readers may refer to (Sen, 2010a). In the following, we briefly discuss a few well known secure routing algorithms used in WSNs.

Liu & Ning propose a protocol called μTESLA– *micro version of the time, efficient, streaming, loss-tolerant authentication protocol-* for providing broadcast authentication in WSNs (Liu & Ning, 2003; Liu & Ning, 2004). The protocol introduces an asymmetry through a delayed disclosure of the symmetric keys, leading to an efficient broadcast authentication scheme. To bootstrap a new receiver, the protocol depends on a point-to-point authentication mechanism in which the receiver sends a request message to the base station and the base station replies with a message containing all the necessary parameters. Since the base station needs to unicast the initial parameters to the individual sensor nodes, a long delay is introduced during network bootstrapping in a large network. Liu et al. have proposed a multi-level key chain scheme for broadcast authentication to overcome this problem (Liu & Ning, 2003; Liu & Ning, 2004).

Zhu et al. propose a scheme known as LEAP (Localized Encryption and Authentication Protocol) that is based on construction of a one-way key-chain for one-hop broadcast authentication (Zhu et al., 2003). In this scheme, each node generates a one-way key chain of certain length and transmits the first key of the key chain to each of its neighbor encrypting it with their pair-wise shared keys. Whenever a node sends a message, it attaches the next authenticated key. The authenticated keys are disclosed in reverse order to their generation.

Deng et al. propose an "*intrusion-tolerant routing protocol in wireless sensor networks*" (INSENS) that adopts a routing-based approach to security in WSNs (Deng et al., 2002; Deng et al., 2003). INSENS operates in two phases: (1) route discovery and (2) data forwarding. During the route discovery phase, the base station sends a request message to all the nodes. Each node receiving a request message records the identity of the sender and sends the message to all its immediate neighbors. The nodes respond with their local topology by sending feedback messages. The integrity of the messages is protected using encryption by a shared key mechanism. A malicious node can inflict damage only by not forwarding packets, but the messages are sent through different neighbors; so, it is likely that a message reaches a node by at least one path. Hence, the effect of malicious nodes is not totally eliminated but is restricted to only a few downstream nodes in the worst case. Finally, the base station computes two independent routing paths for each node from the base station and sends the path information to each node. The second phase of data forwarding takes place based on the forwarding tables computed by the base station.

A suite of security protocols called "SPINS" for WSNs have been proposed in (Perrig et al., 2002). SPINS consists of two building blocks: (1) secure network encryption protocol (SNEP) and (2) the µTESLA protocol. While SNEP provides data confidentiality, two-party data authentication, and data freshness for peer-to-peer communication, µTESLA is a broadcast authentication mechanism.

Du et al. investigate the possible use of public key cryptography (Gura et al., 2004; Gaubatz et al., 2004; Wander et al., 2005; Hankerson et al., 2004) in designing secure routing protocols for WSNs (Du et al., 2005b). The proposed scheme avoids expensive signature verification by using a light-weight one-way hash function for public key authentication. However, it requires the hash values to be distributed in the sensor nodes during the pre-distribution phase which leads to a scalability problem in a large-scale network.

Tanachaiwiwat et al. propose a secure routing protocol called "*trusted routing for location aware sensor networks*" (TRANS) that uses a symmetric key cryptographic scheme based on loose-time synchronization mechanism to ensure message confidentiality (Tanachaiwiwat et al., 2003). Papadimitratos et al. have proposed a secure route discovery protocol that guarantees correct topology discovery in a WSN (Papadimitratos & Haas, 2002; Papadimitratos & Haas, 2006). The protocol relies on the use of MAC and an accumulation of the node identities along the route traversed by a message so that a source node can discover the network topology as each node along the route from the source to the destination appends its identity to the message.

### 3.5 Defense against attacks on the transport layer

For defending against the flooding DoS attacks at the transport layer, Aura et al. propose the use of "*client puzzles*" (Aura et al., 2001). In client puzzle-based schemes, each client has to demonstrate its commitment to the connection by solving a puzzle before it can access any resource in a server. Since an attacker does not have infinite resources, it will be impossible for him/her to create new connections fast enough to cause resource starvation on the serving node. A possible defense against desynchornization attacks is to enforce a mandatory requirement of authentication of all packets communicated between the nodes (Wood &

Stankovic, 2002). If the authentication mechanism is secure, an attacker will be unable to inject any spoofed message.

### 3.6 Defense against the Sybil attack

A defense mechanism against the Sybil attack must ensure that a framework is in place that can validate a particular identity is only being held by a given physical node (Newsome et al., 2004). *Random key pre-distribution* techniques (Eschenauer & Gligor, 2002; Chan et al. 2003; Du et al., 2005a) can effectively be used to defend against the Sybil attack. In random key pre-distribution, a random set of keys or key-related information is assigned to each sensor node so that in the key setup phase, each node can discover or compute the common keys shared by it with its neighbors. The common keys are used as shared secret session keys to ensure node-to-node secrecy.

### 3.7 Defense against node replication attack

Parno et al. propose a mechanism for distributed detection of node replication attacks in WSNs (Parno et al., 2005). In their proposition, the authors have presented two algorithms-(i) *randomized multicast* and (ii) *line-selected multicast* – both of which are based on collaborative participation of multiple sensor nodes. The randomized multicast algorithm distributes location information of a node to randomly selected *witnesses* and exploits the *birthday paradox* to detect replicated nodes. The line-selected multicast algorithm is based on *rumor routing* (Braginsky & Estrin, 2002), and it uses network topology-related information to detect node replication. Line-selected multicast has lower communication overhead than randomized multicast.

### 3.8 Defense against the traffic analysis attack

Deng et al. propose a mechanism for defending against traffic analysis attacks in WSN (Deng et al., 2005a). The authors have identified two different classes of traffic analysis attacks: (1) *rate monitoring attack* and (2) *time correlation attack*. In rate monitoring attack, an adversary first monitors the packet sending rate of the nodes in its neighborhood, and then moves closer to the nodes that have a higher packet sending rate. In a time correlation attack, the adversary observes the correlation in sending times between a node and its neighbor node that is assumed to be forwarding the same packet and deduces the path by following each forwarding operation as the packet propagates towards the base station. The defense mechanism proposed by Deng et al is able to defend against both these attacks (Deng et al., 2005a).

### 3.9 Defense against attacks on sensor data privacy

Since protection of privacy of sensitive data in the sensor nodes in WSNs is an important requirement in many applications, several schemes for this purpose have been proposed by the researchers. These schemes can be broadly divided into three categories: (1) anonymity schemes, (2) policy-based schemes, and (3) schemes based on information flooding.

An anonymity scheme depersonalizes the data before it is released from its source. Gruteser et al. present an analysis on the feasibility of anonymizing location information in location-based services in an automotive telematics environment (Gruteser & Grunwald, 2003). Beresford et al. propose various anonymity techniques for an indoor location system based on the Active Bat (Beresford & Stajano, 2003). Sen proposes an efficient and reliable routing protocol for wireless ad hoc and mesh networks for protecting user privacy while providing robust authentication for the users (Sen, 2010b). The scheme is based on the Rivest's *ring signature scheme* (Rivest et al., 2001).

In policy-based defense mechanisms, decisions on access control and authentication are made on the basis of a specified set of privacy policies. Molnar et al. present the concept of *private authentication* and demonstrate its application in the *radio frequency identification* (RFID) domain (Molnar & Wagner, 2004). Duri et al. propose a policy-based framework for protecting sensor information in which a computer inside a car acts as a trusted agent for ensuring location privacy (Duri et al., 2000). Myles et al. describe the architecture of a centralized location server that controls access requests from client applications through a set of *validator* modules based on a set of XML-coded privacy policies (Myles et al., 2003). Hengartner et al. discuss various challenges that arise in designing the specification and implementation of policies that control access to location information, and present a framework of an access control mechanism (Hengartner & Steenkiste, 2003).

Use of information flooding is a popular approach to achieve privacy in communication. Ozturk et al. propose various modifications to WSN routing protocols for protecting the location information of a source node by using randomized data routing and a *phantom traffic generation* mechanism (Ozturk et al., 2004). Phantom flooding entices an attacker away from the real source towards a fake source called the "*phantom source*". Deng et al. address the problem of defending a base station against physical attacks by concealing the geographic location of the base station (Deng et al., 2005b). Xi et al. propose a successful attack on the flooding-based phantom routing approach presented by Ozturk et al (Ozturk et al., 2004) and describe "*greedy random walk*" (GROW) protocol to reduce the chance of an eavesdropper successfully collecting the communicated location information (Xi et al., 2006). Li et al. propose a scheme that provides both content confidentiality and source-location privacy using a two-phase routing mechanism (Li & Ren, 2009). In the first phase of routing, the source node randomly selects an intermediate node in its neighborhood and transmits the data packets to that node before it is routed to a ring node. This phase ensures protection of the source-location privacy. To be written. In the second phase, the data packets are mixed with packets from other sources through a *network mixing ring* (NMR). This phase of routing provides source-location privacy at the network level. In order to provide high level protection to source-location privacy, it is possible to have multiple mixing rings in the routing process. However, use of multiple mixing rings leads to more energy consumption in the sensor nodes.

### 3.10  Secure data aggregation

In a WSN, certain nodes - called the "*aggregators*" - are responsible for carrying out data aggregation operations so as to optimize the utilization of precious bandwidth of the wireless links. If an aggregator node or a sensor node is compromised, it is easy for an adversary to inject false data into the network. In absence of a robust authentication mechanism, an attacker can fool the aggregators into reporting false data to the base station. For securing the aggregation process in WSNs, two broad categories of techniques are generally used: (1) *plaintext-based protocols* and (2) *ciphertext-based protocols*.

The plaintext-based protocols operate on plaintext information while carrying out the aggregation operation. Hu et al. propose a secure aggregation protocol on plaintext data that uses the μTESLA (Hu & Evans, 2003). In the proposition, sensor nodes are organized into a tree in which the internal nodes act as the aggregators. However, the protocol fails if a parent and one of its child nodes are compromised. Chan et al. have presented a "*secure information aggregation*" (SIA) framework for sensor networks (Chan et al., 2007). Cam et al. propose an "*energy-efficient pattern-based data aggregation*" (ESPDA) protocol for WSNs (Cam et al., 2005; Cam et al., 2006). Cam et al. have introduced another scheme – "*secure differential data aggregation*" (SDDA) -- which is based on pattern codes (Cam et al., 2004). SDDA transmits the differential data instead of the raw data and performs data aggregation on the pattern codes that represent the main characteristics of the sensed data. It also employs a *sleep protocol* to coordinate the activation of the sensing units in such a way that only one of the sensor nodes is activated at a given time for sensing operation. Du et al. propose a "*witness-based data aggregation*" (WDA) scheme for WSNs to ensure validation of data fusion results to the base station (Du et al., 2003). Wagner has studied secure data aggregation in WSNs and has proposed a mathematical framework for formally evaluating the strengths of their security (Wagner, 2004).

Secure aggregation of ciphertext data in WSNs is required to preserve the privacy of sensor nodes in many applications (Acharya et al., 2005; Castelluccia et al., 2009; Girao et al., 2005; He et al., 2007; Westhoff et al., 2006). As a key approach to fulfilling this requirement, "*concealed data aggregation*" (CDA) schemes are proposed in which multiple source nodes send encrypted data to a sink along a convergecast tree with aggregation of ciphertext being performed over the route (Acharya et al., 2005; Castelluccia et al., 2009; Girao et al., 2005; Westhoff et al., 2006; Peter et al., 2010). Two ciphertext-based secure data aggregation schemes are proposed by Castelluccia et al. (Castelluccia et al., 2009) and Girao et al. (Girao et al., 2005). The propositions are based on a particular encryption transformation called "*privacy homomorphism*" (PH). A PH is an encryption transformation that allows direct computation on encrypted data.

### 3.11  Defense against physical attacks on nodes

The sensor nodes in a WSN can be protected against possible tampering by tamper-proofing the physical packages of the sensors (Wood & Stankovic, 2002). Propositions also have been made by researchers for building tamper-resistant hardware in order to make the memory contents on the sensor chips inaccessible to a potential external attacker (Anderson & Kuhn, 1996; Anderson & Kuhn, 1998; Komerling & Kuhn, 1999). Deng et al. propose various approaches for protecting sensors by deploying components outside them (Deng et al., 2005b). Sastry et al. present a protocol called "ECHO" that provides for secure and reliable location verification of the sensor nodes in a WSN (Sastry et al., 2003). Deng et al. discuss various defense mechanisms against search-based physical attacks (Deng et al., 2002). Wang et al. present a systematic modeling framework for "blind" physical attacks on WSNs (Wang et al., 2005). Seshadri et al. propose a mechanism called "*software-based attestation for embedded devices*" (SWATT) to detect a sudden and abrupt change in the memory content of a sensor node that indicates the possibility of an attack (Seshadri et al., 2004).

### 3.12 Intrusion detection

An *intrusion detection system* (IDS) monitors a host or a network for suspicious activity patterns that are outside the normal and expected behavior (Wood & Stankovic, 2002). Research on intrusion detection in WSNs is still in its preliminary stage. Current research focuses on how to detect and eliminate injected false information.

Brutch et al. discuss various types of possible attacks against WSNs and propose various architectures for intrusion detection systems (Brutch & Ko, 2003). Zhu et al. propose an *interleaved hop-by-hop* (IHOP) authentication scheme (Zhu et al. 2004b) which can guarantee that the base station will be able to detect any false injected data packets when no more than a certain number of nodes are compromised. Wang et al. propose a scheme to detect whether a node is faulty or malicious with the collaboration of its neighbor nodes (Wang et al., 2003). Albers et al. present an intrusion detection architecture based on a *local IDS* (LIDS) on each node in a wireless ad hoc network (Albers et al., 2002). Sen proposes an intrusion detection architecture for an ad hoc network in detection activities are carried out locally in each cluster (Sen, 2010d).

### 3.13 Trust management

A popular approach to enforce a high-level of security in WSNs is to deploy trust- and reputation-based frameworks. Issues such as judging the quality and reliability of the sensor nodes and the wireless links, robustness of the data aggregation operation, correctness of the aggregator nodes, and timeliness in packet forwarding by the sensor nodes can be addressed very effectively with the help of trust-based systems. A comprehensive discussion on trust and reputation and various security mechanisms based on these concepts is given in (Sen, 2010c).

Pirzada et al. propose an approach for building trust relationship between the nodes in an ad hoc network based on their packet forwarding behavior (Pirzada & McDonald, 2004). Oram describes various methods of finding paths from a source node to a designated target node in a peer-to-peer computing paradigm (Oram, 2001). Extending this approach, Zhu et al. (Zhu et al., 2004a) provides a practical approach for computing trust in wireless networks.

Sen proposes a trust-based secure and efficient searching scheme for peer-to-peer networks that utilizes topology adaptation by the trusted nodes (Sen, 2011). Yan et al. discuss a trust-based security framework to ensure data protection and secure routing in an ad hoc network (Yan et al., 2003). Ren et al. present a probabilistic approach to model a distributed trust framework for a large-scale ad hoc network (Ren et al., 2004).

Ganeriwal et al. propose a reputation-based framework for high-integrity sensor networks using the beta distribution for reputation representation, updates and integration (Ganeriwal

& Srivastava, 2004). Liang et al. present various models for evaluating the robustness of various aggregation algorithms which can be adapted for WSNs (Liang & Shi, 2005; Liang & Shi, 2008).

## 4. Security Vulnerabilities in CWSNs

In conventional WSNs, the transmission parameters can be changed and the radio frequency (RF) bands can be used in the limits which have been defined by pre-defined standards and spectrum regulations. Based on these specifications, the hardware and the firmware are implemented and they cannot be changed dynamically during the network communications. A *cognitive wireless sensor network* (CWSN), other hand, can communicate in a wide range of spectrum bands by changing its transmission parameters dynamically during network communication in response to the changes in the sensed radio spectrum environment and the signals received from other sensor nodes. This capability is gainfully utilized to realize innovative spectrum management approaches like *dynamic spectrum access* (DSA) in which the allocation of spectrum bands to communication services can change with time or space.

In most of the centralized and distributed approaches of DSA in CWNs, it is assumed that the participating nodes are altruistic and make logical decisions to optimize the use of the spectrum resources. However, such approaches make CWSNs vulnerable to security threats, where malicious cognitive radio nodes may exhibit selfish behavior or would disrupt the communication protocols and algorithms which are designed for optimal spectrum utilization. Identification of various possible attacks on CWSNs is critical so that the networks can be defended against those identified attacks by implementing appropriate security mechanisms. Masquerading is a very common attack on CWSNs in which a malicious cognitive radio node provides false information for the cognitive radio functions such as spectrum sensing or spectrum sharing. The malicious node can also inject false information on the spectrum environment into other CR nodes with the objective of gaining an unfair advantage or just disrupting the CWSN. This type of threat can affect both centralized and distributed CWSNs.

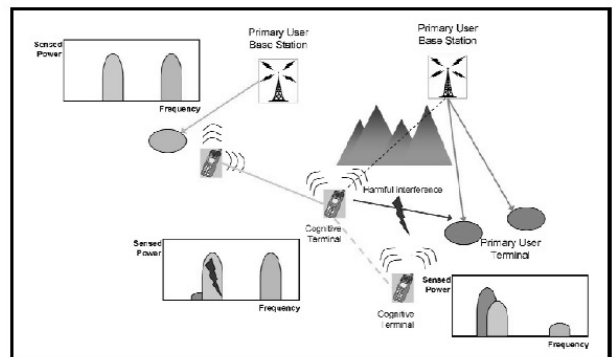

**Figure 2.** Hidden node problem in a cognitive radio network

A node may not always distribute incorrect or incomplete information about the spectrum environment with a malicious intention only. As shown in Figure 2, in case of the hidden node problem, two cognitive nodes may have a different perception of the spectrum because they are located in two different locations and they detect different radio spectrum information.

As in traditional WSNs, *jamming* is the most common DoS attack in CWSNs which can be used to disrupt the communications in a specific spectrum band, or make the management channels of the CWSN ineffective so that cognitive radio related messages cannot be distributed in the network. In addition to all the vulnerabilities in the traditional WSNs that we have discussed in Section 2 of this paper, CWSNs have many other security problems. Some of the attacks which could severely affect the operations in a CWSN are: (i) attacks on the communication protocols, (ii) masquerading attacks, (iii) unauthorized access to the spectrum, (iv) physical attacks on the sensor nodes, (v) internal failures of the sensor nodes, (vi) power exhaustion attacks on the sensor nodes, (vii) attacks on the objective functions of the cognitive engine, (viii) attacks on the administrative policies of the sensor nodes, (ix) attacks on the cryptographic protocols and security schemes implemented in the sensor nodes, and (x) attacks on the privacy of the sensor data.

In the following, we briefly discuss the aforementioned threats and vulnerabilities in CWSNs.

### 4.1 Attacks on the communication protocols

The intention of these attacks is to disrupt the communication in a CWSN. Attacks under this category are of various types such as: (i) replay attack, (ii) denial of service (DoS) attack, (iii) malicious alteration of the cognitive messages, (iv) Sybil attack, (v) hidden node problem, (vi) saturation of the cognitive control channels, (vii) eavesdropping of cognitive radio messages, (viii) disruption of the MAC, network layer, and cognitive engine of the cognitive radio network.

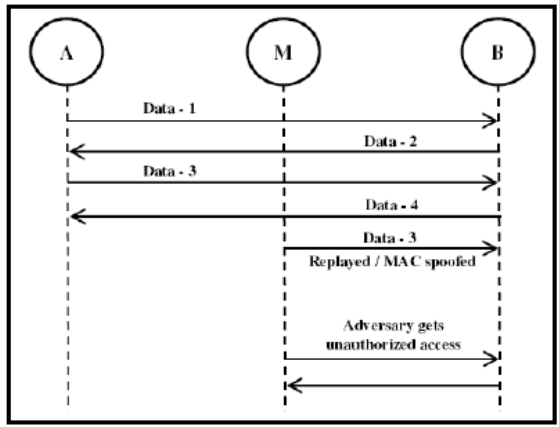

**Figure 3.** An Illustration of *MAC spoofing* and a subsequent *replay attack* launched by a malicious node *M*

In a *replay attack* (Raymond et al., 2007), the attacker replays messages from earlier sessions of communications in the network. This attack is illustrated in Figure 3. The attacker may also send the replayed messages to another node which is not the intended recipient of the message. The receiver of the message, on finding that it is not the intended recipient forwards the messages further so that the message ultimately reaches the actual destination node. However, the delayed messages can lead to spreading of false information, since based on this delay, various characteristics of the network such as channel quality, network topology, routing etc, are computed. Since the nodes in a CWSN share extensive information among each other about various aspects of the network, spreading of false information can cause more damage in a CWSN than in a traditional WSN. For example, if the packets from the *primary users* (PUs) in a CWSN are replayed, the *secondary users* (SUs) might have a wrong perspective of the spectrum as well. This will forbid the SUs from using the frequencies and the protocols used by the attacker resulting in a sub-optimal and inefficient use of network resources.

In a DoS attack, the attacker makes the resources in the network unavailable to its legitimate users. There are many different ways in which a DoS attack may be launched – (i) jamming attack, (ii) collision attack, (iii) routing disruption attack, (iv) flooding attack etc.

In a *jamming attack*, the attacker transmits radio signals that interfere with the radio frequencies used by the nodes in a network. As discussed in Section 2, various ways of launching jamming attacks in WSNs and their defense mechanisms have been extensively studied by researchers over the last decade (Sun et al., 2007). In CWSNs, jamming attacks could be detrimental since it can rapidly exhaust the energies in the nodes and disrupt communication in the network. In a typical jamming attack in CWSN, a malicious node transmits signals at a high power using the PU frequency thereby disrupting the communication in the network. Jamming of the channels used to distribute cognitive messages in CWSNs is another serious threat. This attack can be launched against an out-of-bound *cognitive control channel* (CCC) or in-band CCC if the frequency of the channel is known.

The objective of the *collision attack* is to violate the communication protocols used in CWSNs. While an attacker need not spend much energy in launching such an attack, the attack can cause serious damage in network services. Since the wireless medium is inherently broadcast in nature, detection of collision attacks and identification of the malicious nodes are non-trivial tasks. Since in CWSNs, the SUs share the spectrum, collision attack can easily and very effectively disrupt communications among the SUs. Hence, the collision attacks are more detrimental in CWSNs than in WSNs. In routing *disruption attack*, a malicious attacker does not forward the routing messages. The *grayhole* and the *blackhole* attacks are examples of these types of attacks. As already discussed in Section 2, these attacks are also possible in traditional WSNs. While in a grayhole attack, the attacker selectively drops routing messages, a blackhole node drops all routing packets arriving at it. These attacks not only cause serious disruptions in network communication, but also the spreading of routing misinformation may lead to network partitioning. In a *flooding attack*, a malicious node sends a number of fake connection requests to a target victim node resulting in resource depletion in the latter.

In *malicious alteration of cognitive message attack*, the adversary intentionally changes the cognitive messages in the network so that correct information cannot be exchanged among the nodes.

The *Sybil attack* is launched by an attacker node that can assume multiple identities. This type of attacks can cause routing disruption, and unfair resource allocation in a resource sharing environment and in voting and reputation-based systems. For instance, the Sybil attack may be launched by a malicious node to generate additional reputations for malicious nodes or to change the information about the sensed spectrum.

The *hidden node problem* arises when a CR node is in the protection region of an incumbent node but it fails to detect the existence of the incumbent. For example, if a CR node does not sense the presence of a primary user *base station* (BS) because of an obstacle, it transmits in the same frequency bands of the primary user, causing harmful interference. Depending on their position, other CR terminals sense a different environment, and they can provide additional information to mitigate the threat.

In the *saturation of the cognitive control channel attack*, the attacker launches a DoS attack against the *cognitive control channel* (CCC) by saturation – a large number of cognitive message are sent to the CCC to deny its service in the CWSN. However, some specific designs of the CCC may prevent this type of attack.

In the *eavesdropping of cognitive radio messages*, the attacker passively listens to the cognitive messages, and subsequently uses the information contained on those messages to launch powerful attacks.

In the attack involving *disruption of the MAC, network layer, and the cognitive engine of a CWSN*, the adversary attempts to target the protocols in the higher layers of the stack.

### 4.2 Masquerading attacks

This vulnerability involves the scenario in which a malicious adversary masquerades a primary user in a CWSN. The malicious attacker may mimic the primary user characteristics in a specific frequency band so that the legitimate secondary users erroneously identify the attacker as an incumbent and they avoid using that frequency band. This can be a selfish attack, because the attacker may subsequently use the frequency bands or launch a DoS attack to deny access to the spectrum resources to other secondary nodes in the CWSN. An example of the masquerading of a primary user in a CWSN is shown in Figure 4. A malicious CR node transmits a signal which is very similar to the primary user. On sensing this false signal, other CR nodes detect the presence of an additional primary user, and they avoid using the spectrum bands. In another form of a masquerading attack, a malicious CR node masquerades an honest node while collaborating with the other nodes in a CWSN to carry out important network functionalities such as: spectrum sensing, spectrum sharing, spectrum management, and handling of spectrum mobility. This form of an attack can be dangerous since the malicious node may spread false information about spectrum sharing while participating in the collaborative decision making processes in a CWSN.

One type of masquerading attack – *primary user emulation* (PUE) attack – was first introduced in (Chen & Park, 2006; Chen et al., 2008c). This form of attack is extremely effective in *dynamic spectrum access* (DSA) environments. In a network that allows for DSA, the primary users own licenses to different frequency bands and can use those bands whenever they wish. However, when the primary users are idle, the secondary devices can opportunistically use the spectrum on those bands. Such secondary users need spectrum sensing algorithms to detect when the primary user is active. To attack a DSA algorithm, an attacker needs to create a waveform that is sufficiently similar to that of the primary user so that a false positive may be trigger in the spectrum sensing algorithm. With the false signal being emitted by the attacker, the secondary users in the communication range of a primary user will erroneously conclude that the primary user is active and will cause the system to vacate the channel. As a consequence, the adversary will gain unrivaled access to the specified frequency band. However, this attack is transient in nature since it is a *sensory manipulation attack* (Clancy & Goergen, 2008). Once the attacker vacates the frequency band, the secondary users can resume using the band.

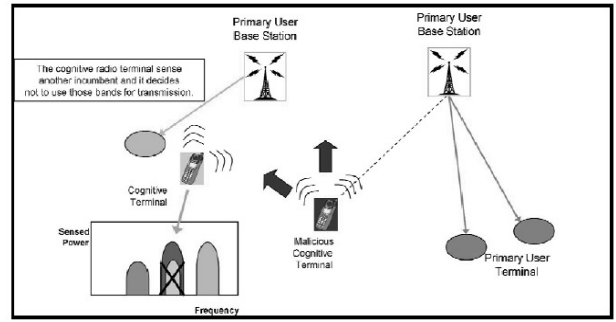

**Figure 4.** An Illustration of a *masquerading attack* on a primary user in a cognitive radio network

Leon et al. have shown how a more sophisticated form of PUE attack can be launched when the attacker has some prior knowledge about the CR network (Leon et al., 2010). This attack can be launched in a CWSN as well. For example, if the attacker knows the exact time of occurrence of the "quite periods" of the CWSN, the attacker can launch the PUE attack during those time periods. A quite period in a CWSN is the interval of time during which all secondary users refrain from transmitting in order to enable a collaborative spectrum sensing environment. If any user receives a signal strength that is beyond a certain threshold it assumes that the signal is emitted by a primary user. Hence, if a malicious user transmits during the quite period, the other nodes wrongly assume it to be a primary user and a PUE attack is successfully launched. Another form of PUE attack can be launched by a malicious adversary whenever a CR network makes a frequent channel switchover. This type of PUE attack can degrade the throughput and may lead to a partial DoS attack.

Other DSA algorithms are more stateful and gather more detailed statistics about the primary users. For example, some DSA algorithms gather channel access information of the primary users and accordingly makes prediction when the channel will be idle based on an estimation algorithm using the past and the current behavior of the primary user (Clancy & Walker, 2006). In such cases, spoofing primary user waveforms can affect the long-term behavior of a secondary user, turning this attack into a belief-manipulation attack (Clancy & Goergen, 2008).

If an adversary attempts to prevent the secondary users from accessing the spectrum bands in a *time-division multiple*

*access* (TDMA) type primary user, the attacker needs to make the channel access pattern by the primary user look random during the learning phase of the secondary users' cognitive engine. As a result, the secondary users are not able to accurately estimate the time when to transmit without interfering with the transmissions from other nodes. This failure in estimation leads to significant decrease in the overall capacity of the network since the spectrum usage is suboptimal.

### 4.3 Unauthorized access of spectrum

A malicious adversary node in a CWSN can launch an attack so that it can use spectrum bands for which it is not authorized or licensed and thereby gain more traffic capacity or bandwidth. Moreover, a malicious node can also emit power in unauthorized spectrum bands to cause DoS to the primary users.

### 4.4 Physical attacks on the sensor nodes

Physical attacks such as tampering with or damaging the hardware of even a very few sensor nodes in a CWSN can have a catastrophic effect on the overall operations of the network. Since the network operations in a CWSN is dependent on the correctness of the critical information exchanged among the nodes, the adverse impact due to node compromise is more severe in CWSNs than in traditional WSNs. In a traditional WSN, failure or compromise of a few sensor nodes usually have a minor impact on the network performance since the connectivity of the nodes are still maintained due to enough redundancy in the network. However, successful operation of a CWSN is dependent on the distributed information among the nodes and their cooperative behavior. Hence, even a single compromised or captured node in a CWSN can be a very powerful weapon in the hand of an attacker for causing disruption in network communication. A compromised node can easily allow extraction of cryptographic keys and modification in the internal code inside the node by an adversary leading to a catastrophic consequence.

### 4.5 Internal failure of the sensor nodes

Failure of CR nodes in a CWSN may occur due to various reasons, i.e., memory fault, physical failure or other hardware failure. The impact of these failures may be quite damaging on the overall network services. For example, a malfunctioning CR node may transmit signals in a wrong frequency band or may not properly participate in important spectrum management-related collaborative decision making.

### 4.6 Power exhaustion attack on the sensor nodes

The sensor nodes are battery powered and energy constrained. To deplete the energies in the sensor nodes, an attacker can launch various types of power exhaustion attacks. For example, the attacker can inflict sleep deprivation attack by engaging in it exchanging of unnecessary message communications to quickly drain off its energy. If the attacker intelligently selects the target nodes, the failure of the nodes can cause network partitioning leading to complete disruption of network operations. In another form of power exhaustion attack, the attacker node can request a channel change very frequently causing a high rate of power usage in the target nodes.

### 4.7 Attacks on objective functions of cognitive engine

In CWSNs, the cognitive engine in a sensor node has many radio parameters under its control. The cognitive engine determines the suitable values of these parameters over time in order to optimize its multi-goal objective functions (Clancy & Goergen, 2008). Various attacks are possible on the learning algorithms of the cognitive engines so that these algorithms produce suboptimal outputs. Since these attacks are targeted on the learning algorithms, they are also known as the *belief- manipulation attacks*. Clancy & Goergen have identified various input parameters of the cognitive engines such as: center frequency, bandwidth, transmit power, type of modulation, coding rate, channel access protocol, encryption algorithm, frame size etc (Clancy & Goergen, 2008). The cognitive radio may have three goals such as achieving *low-transmit power*, *high rate of transmission*, and *high security in communication*. Based on the application currently under use, the cognitive engine assigns different weights to these three goals to maximize its overall objective function. In order to build a robust framework, in the learning phase, the radio tries out various combinations of different values of the input parameters, measures the observed statistics of the network such as bit error rate, and then evaluates the objective function for optimization. Among the three goals, low transmit power and high security in communication are directly controlled by the input. The other goal- high transmission rate is, however, defined by the system output. The adversary can affect the channel in such a way that high-rate in communication is never achieved even when the correct values of the input parameters are chosen (Clancy & Goergen, 2008).

### 4.8 Attacks on the security policies of the sensor nodes

The operating behavior of the sensor nodes in a CWSN are controlled by setting different policies in the nodes. These policies include security and privacy policies that determine access control, authentication, encryption/decryption, key revocation and other related operations. Several attacks may be launched by malicious attackers on these policies such as: (i) *excuse attack* (Araujo et al., 2012), (ii) *newbie-picking attack* (Araujo et al., 2012) etc.

*Excuse attack*: if the network policy and the security policies in the nodes are very generous to allow faster recovery of nodes that might have crashed or damaged and if these policies do not require the nodes to prove their authenticity, a malicious node may exploit these policies by repeatedly claiming to have crashed/damaged. In this way, wrong spectrum information can be sent to the network very often to cause overload in the network leading to partial or complete disruption in network communications.

*Newbie-picking attack*: if a CWSN requires that new nodes pay their dues by making it mandatory for them to give information to the network for some period of time before they can consume any shared resource, a veteran node could move from one newbie node to another, leeching their information without being required to give any information back.

The operating policies in a CR node in a CWSN can be maliciously changed to alter the behavior of the node so that it can be used to support other attacks and threats such as causing harmful wireless interferences to primary or secondary user nodes which may lead to disruption in network operations.

### 4.9  Attacks on the cryptographic protocols

These attacks attempt to break the security mechanisms in the network and the nodes by compromising the cryptographic protocols used. Although, this attack can be launched in various different ways, the ultimate objectives of different forms of attacks are the same: to break the cryptographic algorithm, extract the keys used in encryption, decryption and hash computation, and to identify any possible vulnerability in the software and hardware of the nodes. Since the nodes in a CWSN are inherently resource constrained, the cryptographic schemes implemented in these nodes are light-weight in nature. These light-weight schemes sometimes prove inadequate against powerful and sophisticated attacks launched by high-end automated tools used by the attackers. An attacker may also launch attacks on the key management scheme used in a CWSN by using different strategies such as: naïve brute force attack, sophisticated dictionary attack, and passive session monitoring attack to capture important session-related information. One example of a sophisticated attack is the *differential power analysis* (DPA) attack, in which an attacker measures the strengths of the electromagnetic signals emitted from a target node to successfully identify the key used for encryption and decryption of messages.

### 4.10  Attacks on the privacy of sensor data

Attacks on sensor data privacy are critical attacks since in many deployments of traditional WSNs and CWSNs, the sensor nodes collect and transmit sensitive data which need privacy protection. In CWSNs, the nodes share resources (i.e., spectrum) to establish communications among them and for developing a framework so that they are aware of the environmental parameters under which they are operate. If the privacy of such information is not protected, an adversary can successfully extract sensitive information from several nodes and may launch more powerful attacks on the network using the extracted information.

The attacks on the privacy of the sensor nodes may involve different strategies such as: *eavesdropping*, *impersonation*, and *traffic analysis*. In passive eavesdropping attack, the attacker silently listens to the communications among the nodes to extract useful information about the session, and uses that information to launch a *replay attack* or an *impersonation attack*. In an impersonation attack, the attacker impersonates a legitimate node in the network and establishes communications with other nodes by providing its fake identity. In this way, the attacker can extract the secret cryptographic key used for encrypting the messages. In replay attack, the attacker reuses the captured sensitive information in an earlier session between two legitimate nodes and gains unauthorized access to network resources. An adversary node may monitor the messages to and from the legitimate nodes in a CWSN to launch a traffic analysis attack to deduce the context information of the nodes. The acquired information from traffic analysis is usually used later by the malicious adversary for devising more catastrophic attacks on the sensor nodes and the overall network. For example, spectrum information can be used by a malicious node to identify the weakest spectrum zone and to locate the zone from where the primary users emit their signals.

Location privacy threats represent a unique challenge in CWSN deployment. This is mainly due to the fact that a secondary user's spectrum sensing report on the signal propagation of the primary users are highly correlated to its physical location. Hence, similar to geo-locating individuals via WiFi or Bluetooth signals, a malicious attacker may exploit the correlation to geo-locate the secondary user and thus compromise the user's location privacy. In (Gao et al., 2012), the authors have identified the following location privacy attacks in CR networks which are relevant in CWSNs as well.

*External CR report and location correlation attack*: Since the wireless communication is broadcast in nature, an external attacker may easily get an access to the CR reports of a specific sensing node by eavesdropping and compromise its location privacy by correlating the CR reports and the node's physical location.

*Internal CR report and location correlation attack*: A malicious attacker may participate in the collaborative sensing activities as a legitimate node and then may receive sensing reports from other nodes as rewards. After obtaining the sensing reports, it compromises any of these nodes' location privacy by correlating the node's CR reports and physical location.

*Internal differential CR report and location correlation attack:* Unlike the two aforementioned attacks that are based on individual sensing reports, this attack analyzes the aggregation results of the sensing reports. The adversary appears as an internal node. It estimates a specific node's sensing reports and infers its location information by comparing the aggregation result before and after the node joins/ leaves the network.

The authors have named the first two types of attacks collectively as CR *report and location correlation* (RLC) attack and the third type of attack as the *differential CR report and location correlation* (DLC) attack. To launch the RLC or DLC attack, an attacker normally needs to generate the signal propagation patterns by collecting the average RSS value of each channel at every position.

## 5.  Work on Identification of CWSN Threats

In this section, we present a detailed discussion on the some of the existing works in the literature identifying various types of specific attacks on CR networks and the ways in which these attacks are launched.

### 5.1  Jamming attack

Sampath et al. have discussed various ways in which jamming attacks can be launched on single channel and multi-channel 802.11 standard-compliant network using a single cognitive radio (Sampath et al., 2007). In the single channel jamming attack, the attacker continuously transmits high-power signals in the channel and causes interference with any communication form legitimate users in the network. In order to minimize energy consumption and detection of the attack more difficult, the attacker can also take a periodic jamming strategy in which the attacker transmits jamming packets at periodic intervals of time. In this strategy, the impact of jamming depends on the length of inter-jamming interval, the size of the jamming packets, and the size of the data packets sent to the victim node. It has been found that the impact of jamming degrades gracefully with the increase in inter-jamming interval, while the use of large packet size at the victim node increases the impact of jamming. The authors have argued that there is an inherent trade-off between the network throughput and its attack resistance since the use of small packet size reduces the transmission efficiency while larger packets make the system more vulnerable to jamming attack. The single-channel jamming attack can be made ineffective if the users switch to different channels on observing high packet loss in a given channel. Alternatively, a random hopping across different channels may be done with periodic synchronization to set up the communication links. In multi-channel jamming attacks, the authors have shown how an attacker can manipulate a cognitive radio to switch frequently across different channels and jam multiple channels simultaneously. Since, in addition to fast channel switching, the nodes in a CWSN have advanced channel sensing capabilities, the attacker can use a CR node to build up channel usage patterns of network users, and switch only among the channels which are currently under use. These types of highly intelligent and efficient attacks are very difficult to detect in CR networks (Xu et al., 2005). In multi-channel jamming attacks, it has been shown that with the same energy consumption level, the effective number of channels which are successfully jammed increases with the total number of available channels in the system. The impact of jamming attacks under different radio settings is studied using the network simulator Qualnet. In the simulations, the authors have also examined the difference between the UDP and TCP traffic under jamming attack and the impact of the attack on packet size, and the channel switching delay.

Burbank et al. have presented a detailed description on how various types of jamming attacks can be targeted in a CR network and how adverse these attacks can be on the overall network performance (Burbank et al., 2008). All these attacks are relevant in CWSNs as well. The authors have identified four goals of an attacker: (i) to launch an immediate DoS attack on CR nodes, (ii) to cause degradation in network performance, (iii) to extract important network information for launching more powerful attacks, and (iv) herding – to drive the victim CR network to a state from which a more powerful attack on the network can be launched. As an example, an attacker can launch jamming attack on a CR network to force the network to select an alternative frequency band for the *cognitive control channel* (CCC), wherein another malicious node can eavesdrop on the cognitive messages exchanged in that band.

Sethi and Brown have presented a detailed discussion on various DoS attacks and a framework to analyze those attacks (Sethi & Brown, 2008). The framework, known as the "*Hammer Model Framework*", graphically presents the potential risks sequences for DoS attacks, and investigates various types of vulnerabilities that may prevent CR communication in specific spectrum bands or completely deny a CR network to communicate or induce it to cause harmful interference to its existing legitimate users. In addition to jamming attacks, the authors have also considered attacks related to malicious alterations of cognitive messages and masquerading of a CR node by a malicious adversary. The authors have considered different architectures – collaborative, non-cooperative, centralized, and distributed – to identify, analyze and assess the risk levels posed by the attacks on these different CR design paradigms. The analyses presented in the paper show that while the non-cooperative CR design strategy is not a good idea since it is most vulnerable to attacks, the distributed and cooperative architecture is the most robust design and least susceptible to malicious attacks.

Zhang et al. have classified various vulnerabilities on a CR network based on the CR functions and that adversely affects its learning ability and its ability to gainfully utilize the benefits of its dynamic spectrum access capability (Zhang et al., 2008). Arkoulis et al. have identified, analyzed and explained the security weaknesses and vulnerabilities of cooperative, dynamic and open spectrum access environments that can be targeted by a malicious adversary to disrupt the network operations or degrade its performance (Arkoulis et al., 2008). The authors have followed an approach for identifying threats based on the types of anomalous behavior of the nodes such as: misbehavior, selfishness, cheating, and malicious intention. After identifying the threats, the authors have presented a detailed classification of the twenty two different attacks based on: (i) the attack type, (ii) the type of protocols (distributed or centralized) the attacks target, and (iii) the architecture that these attacks apply for their operations.

Burbank presents some major vulnerabilities in CR networks in general and identifies various challenges in defending against these threats (Burbank, 2008). In order to identify specific security challenges in CR networks which are applicable for CWSNs as well, the author has first pointed out two fundamental differences between a traditional wireless network and a CR network. In the CR networks the attacker has: (i) the potential far reach and long-lasting nature of an attack, and (ii) the ability to have a profound effect on network performance and behavior through simple spectral manipulation by generating false signals. In a CR network, the nodes exchange locally-collected information to construct a perceived environment that that determines the current and future behavior of the nodes. The author argues that in a CR network, a malicious adversary can propagate its behavior through the network in the same way a malicious worm propagates in a network. The adversary can carry out spectral manipulation for influencing the behavior of a set of local CRs or a distant CR as well. The author has also identified various features of CR networks and the implications of these features on potential attacks on these networks. For defending against these attacks, four essential abilities of CR network have been identified: (i) the ability to provide strong authentication to the local observations that are used to form the perceived environment, (ii) the ability to provide a robust and secure framework for exchanging messages among the CR elements, (iii) the ability to authenticate and provide integrity protection to the information exchanged between the CR elements, and (iv) the ability to perform self-analysis of the network behavior.

Brown and Sethi present a multidimensional analysis and assessment of various DoS attacks on all types of CR networks (Brown & Sethi, 2008). The authors have carried out vulnerability analysis of CR network against various DoS attacks using different parameters such as network architecture employed, the spectrum access technique used, and the spectrum awareness model. Three classes of network architecture are considered- (i) non-cooperative, (ii) cooperative and centralized, and (iii) cooperative and distributed. For spectrum access methods used, the CR networks and hence the CWSNs are assumed to operate either in an overlay or an underlay network. In an overlay network, CR searches for white space bands for communication purpose. In the underlay network, the CR uses spread spectrum or ultra-wideband techniques along with transmit power control to minimize interference.

The CR is also assumed to be aware of usage and availability of spectrum in its vicinity using three approaches: (i) geo-location/database approach, (ii) beacon/control signal approach, and (iii) detection/sensing approach. For analyzing various DoS attacks, the authors have categorized these attacks into two types – denial attacks and induce attacks. While the denial attacks are intended to prevent communications in the network, the induce class of vulnerabilities stimulate the CR node to communicate causing interference with a licensed transmitter.

The adverse impact of the above-mentioned attacks is not reflected immediately. However, these attacks cause permission policies to be tightened or eliminated potentially denying network services over a long-term. The authors have shown that the induce attacks may be manifested in five different forms such as: (i) the licensed spectrum appears unoccupied, (ii) the policy is incorrect, (iii) the location information is incorrect, (iv) the sensor provides incorrect measurements, and (v) the commands to the Tx/Rx are incorrect. For each category of attack, the authors have presented detailed discussions on its relative effectiveness and its possible protection measures. In multi-dimensional analysis of DoS attacks, the authors have enumerated a number of metrics for assessing the attack effectiveness such as: *jamming gain* (Brown et al., 2006), *jamming efficiency* (Brown et al., 2006), *packet send ratio* (Xu et al., 2005), and *packet delivery ratio* (Xu et al., 2005). However, no analytical framework is provided for computing the attack effectiveness.

## 5.2 Primary user emulation (PUE) attack

Chen et al. have identified a threat to spectrum sensing, named the *primary user emulation* (PUE) attack in which an adversary's CR transmits signals whose characteristics emulate those of incumbent signals (Chen et al., 2008c). This attack is particularly easy to launch in a CR networks especially in CWSN due to the highly flexible and software-based air interfaces of CR sensor nodes. The study carried out by the authors have shown that such an attack can be catastrophic since it severely interferes with the spectrum sensing process and reduces the channel resources available to the legitimate unlicensed users in the network. The authors have classified the PUE attacks into two categories: selfish attacks and malicious attacks. In a selfish PUE attack, the attacker's objective is to maximize its own spectrum usage. When a selfish PUE attacker detects an unused spectrum band, it transmits a signal that emulates the characteristics of the signals of the primary users thereby preventing other secondary users from competing for the vacant spectrum band. This type of selfish attack is launched by a pair of secondary users whose intention is to establish a dedicated link between them. The objective of a malicious PUE attack, however, is to thwart the DSA process of the legitimate secondary users and to prevent them from detecting and using fallow licensed spectrum bands. The final goal of the attacker is to carry out a DoS attack on the DSA process in a single or multiple bands. To counter this attack, the authors have proposed a transmitter verification scheme called *LocDof* (localization-based defense) that verifies whether given signal is really that from an incumbent transmitter by estimating its location and observing its signal characteristics. For estimating the location of the signal transmitter, the verification scheme employs a non-interactive localization mechanism that utilizes the services of a wireless sensor network to collect snapshots of received signal strength (RSS) measurements across the CR networks. By averaging out the RSS measurements and identifying the RSS peaks, the transmitter location is estimated. The authors have also presented a detailed security analysis of the proposed localization scheme and evaluated its performance using simulations.

In another work, Chen et al. have discussed two different security threats on CR network which are known as *incumbent emulation* (IE) attack and *spectrum sensing data falsification* (SSDF) attack (Chen et al, 2008b). The IE attack is essentially same as the PUE attack since the primary users are also sometimes referred to as the incumbents. The authors have presented two mechanisms to defend against the IE attack. The first mechanism, known as the *distance ratio test* (DRT) uses RSS measurements obtained from a pair of *location verifiers* (LV) to verify the location of the transmitter (Chen & Park, 2006). Since there is a strong correlation between the length of a wireless link and the RSS, the RSS value at two LVs correlate with their respective distances to the location of the transmitter. The second technique proposed by the authors for defending against the IE attack is known as the *distance difference test* (DDT) (Chen & Park, 2006). This technique relies on the fact that when a signal is transmitted form a single source to two LVs, a relative phase difference is observed when the signal reaches the LVs due to the differential distances from the transmitter. This phase difference can be translated into a time difference which in turn can be converted into a distance difference. This expected difference is compared with the measured difference to determine the authenticity of the incumbent signal. If the two values are found to be sufficiently close, the transmitter is considered to be a legitimate incumbent. The other security threat identified by the authors, i.e., the SSDF attack is carried out by malicious secondary nodes that transmit false spectrum sensing data to other nodes in the CR network. This attack is particularly critical in a CWSN, since sending of false spectrum sensing information to a data collector in the network can cause the data collector to make a wrong spectrum sensing decision resulting in a catastrophic impact on the network performance. The authors have argued that to maintain an acceptable level of accuracy in the event of an SSDF attack, the data fusion technique used in the *distributed spectrum sensing* (DSS) needs to be robust against fraudulent local spectrum sensing results reported by malicious secondary nodes. To effectively defend against SSDF attack, the authors have proposed a two-level defense technique. At the first level, an authentication mechanism should be in place for verifying the authenticity of all local spectrum sensing results sent to the data collector, so that any possible replay attack or false data injection attempted by any external entity can be prevented. In the second level, a data fusion scheme should be deployed that is robust against SDDF attack. In order to achieve this, *sequential probability ratio test* (SPRT)-based data fusion technique may be used that supports a variable number of local spectrum sensing results. The robustness of the data fusion technique can also be enhanced by using a reputation-based scheme into the DSS process.

Wang et al. have argued that one of the major challenges in CR networks is to detect the presence of primary users' transmission, since malicious secondary users can send false spectrum sensing information and mislead the spectrum sensing data fusion process to cause collision, interference and inefficient spectrum usage (Wang et al., 2009c). For example, the secondary users can always falsely report the existence of a primary user so that they can occupy the spectrum for a long time. To detect and defend against such security vulnerability, the authors have proposed a malicious user detection algorithm that computes the suspicious level of secondary users and utilizes the suspicious level to eliminate the malicious users' influence on the primary user detection results. The *receiver operating characteristics* (ROC) curves for the primary user detection algorithm used in simulation have demonstrated that proposed security scheme is highly effective in a collaborative spectrum sensing environment.

Clancy and Khawar have highlighted the need of robust signal classification mechanisms for CR networks so that different types of transmitters can be differentiated in a particular frequency band in order to defend against the PUE attacks (Clancy & Khawar, 2009). In a DSA environment, signal classification techniques typically combine signal processing and pattern matching for enabling the secondary users to reliably authenticate signal of a primary user. The signal classification can be achieved by extracting the salient features of the signal using numerous signal processing techniques and then matching these features to a pattern of a known primary user. These features can range from the spectral shape of the primary users signals to high order cyclostationary features of the signal (Cabric et al., 2004). In their proposition, the authors have followed the approach of using unsupervised learning in feature-based signal classification in a DSA environment since unsupervised learning requires minimal pre-configuration in building cognitive radio systems. The signal classification is done using self-organizing maps and then the authors have presented scenarios in which the output classes of the neural network are manipulated by an attacker so that the attacker signals are misclassified by the system as those of the primary users. The authors have also proposed mechanisms to defend against such types of attacks on a CR network. However, the proposed mechanism for attack identification requires sophisticated signal classification algorithms which will pose significant challenges in their implementation in CWSNs.

Anand et al. have presented a novel analytical framework to analyze the feasibility of PUE attack in a CR network which can be applied to a CWSN as well (Anand et al., 2008). The authors derived mathematical expressions for computing the probability of a successful PUE attack and have also provided the lower bounds on the probability of a successful attack on a secondary user by a set of co-operating malicious users using *Fenton's approximation* (Fenton, 1960) and *Markov inequality* (Ross, 2009). For developing their proposed model, the authors have considered a fading wireless environment with losses due to attenuation, fading and shadowing and have analyzed various parameters that can affect the feasibility of a PUE attack. The analysis shows that the probability of a successful PUE attack increases with the distance between the primary transmitter and the secondary users.

## 5.3 Masquerading attack

Masquerading attack on a CR node and the attack involving malicious alteration of CR nodes for disrupting spectrum sensing functions have attracted considerable research attention.

Wang et al. have studied the adverse effect of malicious and compromised secondary users in a CR network (Wang et al., 2009b). The compromised secondary users can report false spectrum detection results in a collaborative spectrum sensing in a CR network and significantly degrade the network performance. The authors have considered a scenario in which there are multiple number of malicious secondary users in a CR network and have proposed an "onion-peeling approach" to defend against these multiple untrustworthy secondary users. The proposed approach is based on computation of the suspicion level of each secondary node based on its spectrum sensing report. If the suspicion level of a node exceeds a predefined threshold value, the node is considered to be malicious and its report is excluded in the collaborative spectrum sensing decisions. In summary, the proposed mechanism has two distinct advantages: (i) it is self-adaptive and hence does not need to know the number of malicious nodes in the network beforehand, and (ii) it has a better performance compared the similar security schemes that work based on the *a priori* knowledge of the maximum number of compromised nodes in the network. The simulation results have shown that while the presence of malicious secondary node adversely affects the performance of the CR network, the proposed security scheme greatly improves the network performance by efficiently detecting the malicious and compromised nodes. In a similar work, Wang et al. have investigated ways to improve the security in collaborative sensing so that malicious secondary users cannot send false spectrum sensing reports to attack the spectrum-related data fusion process (Wang et al., 2009c). The proposed malicious user detection scheme is based on computation of trust values as well as the consistency values that are subsequently used to detect malicious behavior so that the nodes exhibiting malicious behavior are not allowed to influence the primary user detection results. In this way, malicious secondary users are prevented from masquerading the primary users' signals.

Chen et al. have considered the security issues related to malicious secondary users reporting false spectrum sensing information due to Byzantine failure in a distributed spectrum sensing environment in a CR network (Chen et al., 2008a). The Byzantine failures, such as device malfunction or attacks severely affects the spectrum sensing in a CR network since these failures or attacks can enable an attacker to constantly report the spectrum in a band being in use causing severe under-utilization, or it might cause missed detection of primary users resulting in interference with its communication. To increase the resilience and robustness of CR networks under such attacks, the authors have proposed a scheme called *weighted sequential probability ratio test* (WSPRT). The analysis and simulation results reported by the authors show that the proposed scheme can guarantee the accuracy of sensing results even when a considerable number of secondary users report false sensing information. Hu et al. have also addressed the issue of Byzantine failures of secondary users in a CR network, and have proposed a security mechanism that is similar WSPRT (Hu et al., 2009). In this scheme, the binary local reports used in WSPRT are replaced with *N*-bit local reports to achieve an enhanced detection performance. The proposed scheme also uses three types of reputation rating evaluation schemes: neutral, punitive and heavy punitive. The simulation results have shown that the heavy punitive scheme is the most robust against Byzantine or malicious sensing terminals.

Mody et al. have discussed various security threats in IEEE 802.22 standard-compliant devices which are deployed in CR networks (Mody et al., 2009). The authors have also presented a framework for attack classification for CR networks that includes attacks such as: jamming, malicious alteration of cognitive messages, masquerading of primary users, malicious alteration of CR nodes, and masquerading of CR nodes. This attack classification framework is applicable for CWSNs as well.

### 5.4 False spectrum reports sent by secondary users

The threats due to false spectrum reports sent by malicious secondary users have also received attention from the research community. Two notable works in this area can be found in (Safdar & O'Neill, 2009; Li & Han, 2010).

Safdar and O'Neill have identified the need of securing the cognitive control channels to perform channel negotiations before any actual data transmission among the nodes in a CR network (Safdar & O'Neill, 2009). Securing the control channels ensures that CR messages communicated over these channels cannot be altered by a malicious adversary. This protection is critical in CWSNs which are deployed for mission-critical applications. The authors have proposed a novel framework for providing common control channel security for co-operatively communicating CR nodes so that a pair of CR nodes can authenticate each other prior to any confidential channel negotiations to ensure subsequent security against attacks. The proposed detection approach is based on identifying the malicious secondary nodes that possibly send false reports in collaborative sensing networks.

Li and Han have discussed a critical security issue in collaborative spectrum sensing, in which malicious secondary user(s) sends false spectrum report to thwart the spectrum data fusion process (Li & Han, 2010). The authors observe that while it is mandatory to detect potential attackers and make attack-proof decisions for spectrum sensing, most of the existing attacker detection schemes assume a priori knowledge of the attacker's strategy and thus apply the Bayesian detection of attackers. However, in real-world CR networks, the data centers do not have any prior information about the strategy adopted by the attackers. To overcome this shortcoming of the existing detection algorithms, the authors have proposed an abnormality detection approach that is based on counterpart technique in data mining. The performance of the attacker detection scheme in presence of a single attacker in the network is analyzed explicitly. The single-attacker scenario is considered in two different cases. In the first case, known as the independent attack, the attacker does not know the reports of the honest secondary users. In this case, it is numerically shown that the attacker can certainly be detected as the number of spectrum sensing rounds tends to be very large. In the second case, known as the dependent attack, the attacker knows the reports of all the secondary users, and sends its report based on the information in the reports of the other secondary users. In this case, the authors have shown that the attacker can successfully avoid being detected if he/she has perfect information about the missed detection and false alarm probabilities of the detection system. Finally, the performance of the detection scheme in presence of multiple attackers is analyzed using numerical simulations.

### 5.5 Attacks on cognitive control channels

Prasad has argued that design of CR network poses many new technical challenges in protocol design, power efficiency, spectrum management, spectrum detection, environment awareness, novel distributed algorithms design for decision making, distributed spectrum measurements, quality of service (QoS) guarantees, and security (Prasad, 2008). Overcoming these issues becomes even more challenging due to non-uniform spectrum and other radio resource allocation policies, economic considerations, the inherent transmission impairments of wireless links, and user mobility. In presence of these challenges, ensuring security and robustness in network operations becomes extremely critical. The author have identified various research challenges for security in CR networks and have presented the security and privacy requirements, threat analysis and an integrated framework for security using fast authentication and authorization architecture.

### 5.6 Attacks on the MAC layer

Zhu and Zhou have provided a security analysis of the MAC protocols used in CR networks by investigating the impact of DoS attacks on these protocols (Zhu & Zhou, 2008). The authors have argued that all MAC protocols for a multi-hop CR networks use a common control channel to perform channel negotiation before data transmission. Insecure transmissions of control channels provide opportunities to malicious adversaries for launching DoS attacks. In order to make a security analysis of the MAC protocols, the authors have distinguished two types of attacks and then discussed how DoS attacks can be successfully launched on the MAC protocols. The authors have also presented a detailed discussion on MAC layer greedy behaviors in CR networks and the factors that determines the efficiency of the DoS attacks.

### 5.7 Attacks on the cognitive engines

Clancy and Goergen have defined three classes of attacks on the cognitive engine of CR networks (Clancy & Goergen, 2008). All these types of attacks manipulate the behaviour of the CR system such that the radio acts either sub-optimally or even sometimes maliciously. Three classes of attacks that are identified by the authors are: (i) sensory manipulation attacks against policy radios, (ii) belief manipulation attacks against the learning radios, and (iii) self-propagating behaviour leading to cognitive radio viruses. In a policy radio, the main vulnerability lies in the fact that an attacker can spoof faulty sensor information that can cause the radio to select a sub-optimal configuration. Since the radio sensors take digitized RF and extract useful statistics from it, by manipulating the RF that is available to the radio, an attacker can cause faulty statistics to appear in the CR knowledge base. The learning radios are also vulnerable to the same threats as the policy radios. However, since a leaning radio uses all its past experiences in building its long-term behavior, attacks on it are much more detrimental. For example, an attacker can transmit a jamming signal whenever a policy radio attempts to switch to a faster modulation rate. This will always force the CR to operate at a lower modulation rate, resulting in lower links speeds and link degradation. The authors have called these attacks as belief manipulation attack since these attacks can potentially have much longer-term adverse impact of the learning radios. The self-propagating behaviour of the radio can be utilized by a malicious attacker to launch the most powerful type of attack. In such an attack, the state on radio causes a behaviour that can induce the same state on another radio. Once the target radio attains the state, it exhibits behaviour that leads to a state change in another radio so that it attains the same state. Eventually, the same state propagates through all radios in a particular area in the CR network. The resultant effect is that of a cognitive radio virus that propagates through the network. When acting optimally, all device traverse through the same states and execute the same behavior. However, an attacker can influence this equilibrium in such a way that asymptotic state attained by all the nodes are not optimal (possibly far from optimal) and even malicious.

### 5.8 Threats related to the hidden node problem

The threats related to the hidden node problem in CR networks have also been studied extensively by the researchers. The notable works in this domain are (Biswas et al., 2009; Nuallain, 2008; Bliss, 2010). Biswas et al. have proposed a technique to handle both wideband and cooperative spectrum sensing tasks in a distributed spectrum sensing environment (Biswas et al., 2009). In the proposed approach, the wideband spectrum is divided into several sub-bands and a group of CR nodes is assigned for sensing of a particular narrow sub-band. A cognitive base station is sued for collecting the spectrum sensing results and for making the final decision over the full spectrum. The simulation results have shown that the proposed approach minimizes time and energy spent for wideband spectrum scanning by a CR node, and it also effectively detects the primary users in the wideband spectrum. Nuallain have presented a fast and robust propagation method for addressing the hidden node problem in a CR network (Nuallain, 2008). The proposed method in conjunction with a radio environment mapping server can be used to address the hidden node problem and also to ensure security and reliability in CR networks. The authors have also provided a roadmap for the development of the propagation method so that sufficient accuracy in the results can be achieved. Bliss have investigated the optimal spectral efficiency for a given message size that minimizes the probability of causing disruptive interference for a CR network (Bliss, 2010). The ultimate goal of the work is to have an optimization between longer transmit duration and wider bandwidth versus higher transmit power so as to tackle the hidden node problem in wireless network communication. The probability of interference is assumed to be characterized by the probability that the signal power received by a hidden node in a wireless network exceeds some pre-defined threshold value.

## 6. Security Mechanisms for CWSNs

In this section, we first identify the main security requirements in a CWSN and then discuss various security schemes for defending against attacks in these networks. In a CWSN, the sensor nodes participate in collaborative spectrum sensing activities. The main security requirements in these networks are as follows (Gao et al, 2012):

**Authentication mechanisms:** A robust authentication mechanism is a prime requirement in collaborative spectrum sensing for ensuring that only the legitimate nodes in the network can only access the spectrum. The authentication scheme may have different perspectives to different categories of nodes in a CWSN. The authentication of the primary users is a critical issue since in an attacker may transmit signals with high power that has close resemblance with the signals of a primary user, thereby launching a *primary user emulation* (PUE) attack (Chen et al., 2008c; Liu et al., 2010). To prevent such an attack, the secondary users should have a robust verification scheme for verifying the authenticity of the received signals when they sense the channel. Similarly, when the secondary users receive the sensing reports from other users, they should be able to verify the authenticity of the other secondary users. Otherwise, a potential adversary may be able to spoof the identity of a secondary user for sending false sensing reports. The authentication of sensing reports distributed across the network is also a very important issue. Even if the identities of the secondary user and their authentication are done during the sensing reports aggregation process, it is still possible for a malicious secondary user to send false sensing reports. This attack is known as *spectrum sensing data falsification* (SSDF) attack (Wang et al., 2009b; Fatemieh et al., 2011). It is, therefore, mandatory that each sensing report used in the aggregation process is authenticated as well.

**Incentive mechanisms**: Most of the collaborative sensing schemes are based on a simple assumption that all the secondary nodes voluntarily participate in spectrum sensing. However, this assumption may not hold good for selfish secondary users who may not cooperate in order to conserve their own resources like energy and memory (Song & Zhang, 2009; Wang et al., 2010). Such selfish behavior may seriously degrade the performance of a CWSN. Incentive schemes are necessary for minimizing the probability of such selfish behavior.

**Data and message confidentiality:** The sensing reports need to be well protected so that these messages are not misused by unauthorized external users who may monitor the communication channels by eavesdropping in order to gain useful information. Data and message confidentiality can be achieved by using end-to-end robust encryption algorithms which in turn needs mutual authentication and authorization among the collaborating nodes participating in spectrum sensing.

**Privacy preservation of sensor data:** Privacy protection is primarily for preserving the anonymity of the sensing nodes and/or privacy of its location. Location privacy protection attempts to prevent a possible adversary form linking a sensing node's sensing report to the physical location of the sensing node.

In order to satisfy the aforementioned security requirements and to defend against various possible attacks on the sensor nodes in a CWSN, various defense mechanisms have been proposed by the researchers. In the rest of this section, we present a brief discussion on some of these propositions. Some specific security mechanisms for CWSNs are discussed in detail in Section 6.

The masquerading attacks and the attacks involving distribution of false information in cooperative CR networks for CR-related functions (i.e., spectrum sensing, spectrum management, spectrum sharing, and spectrum mobility) have attracted significant attention of the research community since these attacks are considered to have most adverse impact on the network operations. In majority of the existing security schemes, the secondary users are usually assumed to be trustworthy. However, such schemes will be broken in the event of any masquerading attack launched by a malicious secondary user. A significant number of schemes have been proposed by the researchers for addressing the vulnerabilities and improving the robustness of the collaborative sensing algorithms used in the CR networks in general, and CWSNs in particular.

The security mechanisms for CWNs can be broadly divided into the following categories: (i) security mechanisms for enhancing the robustness in sensor inputs, (ii) security mechanisms based on the reputation and trust of the nodes, (iii) defense schemes based on identification of masquerading attack by signal analysis, (iv) robust authentication schemes using appropriate cryptographic algorithms, (v) security mechanisms for preventing unauthorized access to the spectrum, (vi) security mechanisms for defending against attacks on the MAC layer and the cognitive engine of the network, (vii) protection mechanisms for increasing the robustness of the cognitive control channel against jamming and saturation attacks, and (viii) security mechanisms deployed using geo-location database of the primary users in the network.

In the following, we present a brief discussion on these various types of security mechanisms deployed in CR networks.

### 6.1 Enhancing robustness in sensor inputs

If the reliability of sensor inputs is enhanced, many of the attacks on CR networks can be effectively defended. For example, if the cognitive radios can minutely identify the differences between interference and noise, they can distinguish and hence identify natural and artificial RF events. Such sensors can feed specialized policy engine algorithms that specifically look for hostile signals that may be try to subvert a radio's belief. In a distributed computing scenario, a group of cognitive nodes can fuse sensor data to improve the performance of the overall network. For example, if multiple sensor nodes exchange time-synchronized RF information, they can cross-correlate the exchanged information to arrive at a more precise identification of an attacker in the network. The task becomes challenging, however, since the all sensory inputs are imprecise to a certain extent. Therefore, for each input, the designers should ideally quantify the probability of detection failure in both benign and hostile environments. Since, in certain scenarios, the attackers have power limitations, the designers may get an opportunity to compute the theoretical upper bound of the attack effectiveness and make appropriate risk mitigation.

### 6.2 Reputation- and trust-based security systems

A significant number of schemes are proposed by the researchers using reputation and trust of the nodes in a CR network for defending against attacks such as malicious alteration of cognitive messages, masquerading of the primary users, malicious alteration of cognitive radio nodes, and masquerading of cognitive radio nodes. Using the concepts of reputation and trust, a node can be mapped to a particular level of trustworthiness. On the basis of the information that the node has shared with other nodes regarding its spectrum sensing information, the trust and reputation metrics are computed for a node (Sen, 2010c; Sen, 2013). If the information shared by the node is found to be incorrect after a certain number of iterations by its neighbors, the node is considered to be malicious and appropriate security policies are applied to deal with the identified malicious node.

Zeng et al. have proposed a reputation-based *cooperative spectrum sensing* (CSS) framework using trusted nodes in a CR network (Zeng et al., 2010). The authors categorize the reputation of each node into one of the three categories: (i) *discarded*, (ii) *pending*, and (iii) *reliable*. At first, sensing information from the trusted nodes only is considered reliable and used in the decision making. Reputations of other nodes are put in the pending state, and they are accumulated through a consistency check by the global and local sensing decisions. The information received from the nodes which have their trust values greater than a pre-defined threshold is then considered reliable and their sensing results are incorporated in the CSS. The use of reputation system increases the robustness of the proposed cooperative sensing scheme.

Duan et al. have proposed a reputation-based secure cooperative spectrum sensing algorithm in a CR network (Duan et al., 2009). The proposed algorithm uses a double threshold detector and is effective in mitigating the adverse effects of shadowing and fading in wireless channels in cooperative spectrum sensing and in eliminating the problem related to fail sensing in CR networks. The authors have analytically derived the closed forms for the normalized average number of sensing bits, the probabilities of the detection and false-alarm rates. The simulation results show that the average number of sensing bits in the proposed algorithm decreases to a large extent without fail sensing problem and the sensing performance is improved as compared with the conventional double threshold detection and the conventional single threshold detection.

Several work have been done for detecting malicious nodes using trust-based decision framework in CR networks (Wang et al., 2009b; Wang et al., 2009c; Kaligineedi et al., 2008). While in (Wang et al., 2008b), the attack is considered to be launched by a single malicious node, the authors in (Wang et al., 2009c; Kaligineedi et al., 2008) have presented scenarios in which multiple malicious node launch a cooperative attack in a distributed manner. All these detection schemes, however, use a *centralized spectrum sensing architecture*. The serious drawback of the approach is that the complexity of the detection algorithm becomes prohibitively high in the event of a large number of malicious nodes launching a cooperative and distributed attack. To simplify the problem, the authors have presented an *onion-peeling* mechanism where all the nodes in the CR network are initially considered malicious (till they are proved to be honest) when a specific threshold of abnormality in their behavior is crossed.

Li and Han propose an anomaly detection algorithm for identifying attackers in a collaborative spectrum sensing environment (Li & Han, 2010). The proposed scheme does not assume any *a priori* information about the strategy used by the attackers in launching the attack. Hence, it is robust against most of the spectrum misuse attacks.

Kaligineedi et al. describe an attack detection scheme to identify malicious users that prevents spreading of false spectrum sensing information in a CR network (Kaligineedi et al., 2008). The scheme uses the average power obtained from the real-valued reports received from cognitive nodes for making a global decision on spectrum sensing. The attack detection is done using a *trust factor* mechanism. The authors have considered different kind of malicious behaviors of the nodes such as: (i) *always yes* nodes, (ii) *always no* nodes, (iii) nodes *producing false sensing reports once in a while*. An "always yes node" reports a value above the threshold (i.e., it declares that a primary user is present) all the time. On the other hand, an "always no node" reports a value below the threshold thereby always declaring the absence of primary user in its vicinity. While the "always yes nodes" increase the probability of false alarm, the "always no nodes" decrease the probability of detection. The nodes that produce false values once in a while significantly affect the performance of the sensing system during those intervals of sensing in which they send false information. The proposed scheme can identify each of the three categories of malicious behaviors if the energy values of the malicious nodes differ in distribution from the underlying distribution of the energy values of the legitimate nodes. The malicious nodes are detected using an outlier detection method that assigns a trust factor to each user based on the reliability of the past and the present spectrum sensing reports sent by the user.

Chen et al. present a security scheme based on *weighted sequential probability ratio test* (WSPRT) to deal with the Byzantine failures on nodes in the data fusion process of collaborative spectrum sensing in a CR network (Chen et al., 2008a). In this scheme, each node is allocated a reputation rating based on the consistency of the local sensing report of the node with the final decision in the spectrum sensing.

Peng et al. discuss the motivations for cross-layer design in CR networks and its various security aspects (Peng et al., 2009). The authors also propose a novel architecture in which dynamic channel access is achieved by a cross-layer design between the PHY and the MAC layers in a cognitive node. It has also been shown that among various alternatives, the centralized cooperative security architecture is more efficient and effective for defending against Byzantine failure of nodes.

Anand et al. have analyzed the performance limitations of collaborative spectrum sensing in a DSS environment under Byzantine attacks where malicious users send false spectrum sensing data to the fusion center leading to increased probability of incorrect sensing results and wrong global decisions being taken by the CR (Anand et al., 2010). The authors show that if the percentage of Byzantine attackers in a cognitive network exceeds a certain threshold value, the data fusion scheme utilized for spectrum data fusion becomes highly unreliable. Under such situations, no reputation-based fusion system can achieve any performance gain in the data fusion operation. Further, the authors also present an optimal set of attack strategies for a given set of attack resources and propose possible counter measures at the data fusion center.

Xu et al. present a collaborative sensing algorithm that uses an energy detector with double thresholds and an extended data fusion rules to identify untrusted and possibly malicious CR nodes (Xu et al., 2009). The authors propose the use of an energy detector with two thresholds in a censoring sensor to reduce the transmitted bits in a bandwidth-limited channel. In the censoring sensor, the users whose received energy is above or below the two pre-defined thresholds make local decisions and send one-bit results to the fusion center. The users, whose energy lies in between the pair of threshold values, send no information. The authors also compute the probabilities of detection and false alarms of three different data fusion rules proposed in the scheme. The simulation results have demonstrated the effectiveness of the proposed scheme in defending against untrusted secondary users.

Yu et al. discuss the security issues related to the spectrum sensing data falsification (SSDF) attacks on *cognitive radio-mobile ad hoc networks* (CR-MANETs) in which attacker(s) sends false local spectrum sensing results in a DSS environment (Yu et al., 2009). The authors propose a consensus-based cooperative spectrum sensing scheme for defending against the SSDF attack in CR-MANETs. The proposed scheme is inspired from the self-organizing behavior of animal groups. Unlike most of the schemes for defending SSDF attack, the proposed mechanism does not need a common receiver to carry out data fusion for global decision on spectrum sensing.

## 6.3  Masquerading attack detection by signal analysis

Signal analysis is a popular and widely used technique for identifying malicious attacker(s) in CR networks since this method is very effective in detecting an attacker that masquerades as an incumbent transmitter by transmitting unrecognized signals in one of the licensed bands. Since by transmitting unrecognized signals in the licensed bands the attacker can effectively prevent secondary users in the CR network from accessing the same spectrum band, detection of such attacks is critical. Spectrum sensing can be done in a variety of ways. Some of the commonly used spectrum sensing methods are: sensing based on energy detection (also known as radiometry or periodogram), waveform-based sensing, cyclostationarity-based sensing, radio identification-based sensing, matched filtering, multi-taper spectral estimation, wavelet transform-based estimation, Hough transform, and time-frequency analysis (Yucek & Arslan, 2009). However, each of these spectrum sensing techniques are vulnerable to attacks since an adversary can masquerade a primary or a secondary user by emulating its signal.

Various security schemes have been proposed by researchers to detect and defend against such attacks. Some of the mechanisms are briefly discussed in the following.

Chen and Park propose a security mechanism for defending against masquerading of a primary user by a malicious adversary (Chen & Park, 2006). The proposed scheme is based on a transmitter verification procedure that employs a location verification scheme to distinguish incumbent signals (i.e., signals from a primary user) from unlicensed signals masquerading as incumbent signals. Location verification is carried out using two approaches: (i) *distance ratio test* (DRT) using the *received signal strength indicator* (RSSI) of a signal source and (ii) *distance difference test* (DDT) using the relative phase difference of the received signal as the signal is received at different receivers. The authors assume that the location information of some of the nodes in the network is always known *a priori* since these nodes are either fixed or they use trusted *global positioning system* (GPS) information. These nodes perform DRT and DDT operations within their coverage areas and also serve as the *location verifiers* (LVs). The LVs exchange the location information of incumbent transmitters through a cognitive pilot channel. However, the difference in the radio propagation paths among the LVs which are used in DRT operation can make identification of attackers a very difficult task. The problem is even more critical in urban environments in which buildings and other tall structures frequently cause *multipath fading* in the wireless channels. However, DDT technique does not suffer from this problem and it is usually used to identify masquerading attack in CR networks.

For identifying and isolating malicious users in a CR network, Zhao and Zhao propose a cooperative detection scheme (Zhao & Zhao, 2009). In the proposed scheme, the secondary users collaborate by exchanging and using decision fusion on the local decision results instead of using the detected energy. A mechanism of weighted coefficients is used which updates the weights of the coefficients recursively according to the deviations between separate decision information and the combined final results. The proposed scheme has lower complexity and better performance compared to the existing similar schemes as demonstrated by the simulation results.

Zhao et al. present an identification mechanism of the nodes in a CR network using an analysis of the transmitted signals in which *wavelet transform* is used to magnify the fingerprints of the *transmitter characteristics* (Zhao et al., 2010). This approach is based on PHY-layer authentication and can effectively prevent the PUE attack. The transmitter location fingerprints are extracted from the wireless medium in a multipath propagation environment and a wavelet transform is used to extract the characteristics of the fingerprints. An accurate extraction of signal fingerprints enables a reliable detection of the primary users. However, the radio propagation errors can increase the probability of false alarms. This may make the scheme unreliable in certain wireless environments. Moreover, if an attacker can successfully emulate the transmitter fingerprints, the security of the proposed scheme will be broken.

Afolabi et al. have described a PHY layer attack model that exploits the adaptability and flexibility of CR networks (Afolabi et al., 2009). The authors also propose a *waveform pattern recognition scheme* to identify emitters and detect camouflaging attackers by using *electromagnetic signature* (EMS) of the transceiver. The EMS of a device is computed based on the distinctive behavior in the waveform being emitted by the components of the transceiver including the frequency synthesis systems, modulator sub-systems, and the RF amplifiers. EMS serves as a very reliable and accurate parameter for identification of a node. However, it may be difficult to maintain a database of EMSs of all the devices in a large network and the EMS of a device may change with the aging of the device. Therefore, the scheme may be unreliable in real-world deployment.

Clancy and Khawar present sophisticated signals processing algorithms like *cyclostationary analysis*, *classification engines*, and *signal feature extraction* for identifying false signals in CR networks (Clancy & Khawar, 2009). The authors focus on the use of unsupervised learning in feature-based signal classification, since this technique requires minimal pre-configuration in building cognitive radios. They have shown how self-organizing maps can be used for signal classification and have presented scenarios in which the output classes of the neural network are susceptible for manipulation by an attacker. Due to these possible manipulations, the attacker's signals can be erroneously classified as those of the primary users by other users in the network. In this way, the attacker need not mimic the spectral properties of the primary users, yet its gets unrivaled access to the spectrum. The authors have also provided recommendations to mitigate the impact of the attack on the CR network. However, the proposed detection algorithms are too complex and computation-intensive for resource constrained sensor nodes in CWSNs.

## 6.4 Roust authentication using cryptography

Cryptographic techniques are most commonly used in authentication protocols for wireless networks. However, in CWSNs, authentication mechanism should be adaptable to all communication protocols with which the nodes have to interface. Hence, implementing authentication protocols for CWSNs poses significant design challenges. Over the last few years, this problem has attracted significant attention from the research community.

Kuroda et al. present a radio-independent authentication framework for CR networks that is independent of the underlying radio protocols used and that can be integrated with the *extensible authentication protocol* (EAP) (Kuroda et al., 2007). The re-keying mechanism used in the authentication protocol uses user-specific information, such as location information, as the key seed. The keys used in encryption and authentication are derived from the historical location registry of the mobile device which is securely maintained in a trusted center. The keys are updated frequently based on the change in location of the devices. The authors also evaluate the confidentiality of the key management scheme and its integration related issues with EAP. The protocol is suitable for deployment in real-world networks since it allows fast switchover in CR network and does not need any communication with the *authentication authorization and accounting* (AAA) server for any re-authentication of the CR nodes.

The authentication protocols may have different implementations in different architectures of the CR

networks. While a network having a centralized CR architecture will usually deploy a centralized authentication server for authentication of the nodes and key management-related functions, a distributed CR architecture will be ideally suited for distributed authentication approach, e.g. *threshold authentication*. Capkun et al. propose a set of key management algorithms for a distributed *mobile ad hoc network* (MANET) following the approach in the *pretty good privacy* (PGP) algorithm, in which each node is responsible for creating its public and private keys (Capkun et al., 2003). The proposed scheme is a fully self-organized public key management system that allows users to generate their public-private key pairs, to issue certificates, and to perform authentication even in the event of transient network partitioning and without the presence of any centralized key management entity in the network. The key management scheme is distributed in its true sense since it does not require any trusted authority even at the time of system bootstrapping.

Jakimoski and Subbalakshmi propose an efficient and provably secure protocol that can be used in a CR network to protect the spectrum decision process against a malicious adversary (Jakimoski & Subbalakshmi, 2009). The proposed protocol is to guarantee a secure spectrum decision process in a clustered infrastructure-based network where the spectrum decisions are made at periodic intervals. The decision in each cluster is taken independently of the decisions in other clusters. The protocol is provably secure and it can guarantee that a malicious outsider and a limited number of malicious or selfish insiders cannot make significant adverse impact on the spectrum decision results. The authors have also shown that the proposed protocol is more efficient than the solutions that are based on digital signatures or key establishment protocols.

The CWSNs should also ensure authorization of the cognitive sensor nodes for transmitting specific spectrum bands or for performing specific network functions. The authorization is often conditional to the nature of the spectrum environment, i.e., the presence of primary users in the area. The authorization is needed to define the roles of the CR nodes in performing the CR functions in the network. For both authentication and authorization purposes, the CR nodes exchange authentication information (e.g., the certificates) through a common channel, which is usually the *cognitive control channel* (CCC). Safdar and O'Neill propose a security framework for protecting the information exchanged over the CCC in a CR network (Safdar & O'Neill, 2009).

## 6.5 Prevention of unauthorized spectrum access

Several propositions are made by researchers for defending attacks in which a malicious node accesses spectrum in a CR network and then either uses the spectrum selfishly or launches a DoS attack on the primary users. In the following, we provide a brief discussion on some of these schemes.

Xu et al. present a framework known as TRIESTE (Trusted Radio Infrastructure for Enforcing SpecTrum Etiquettes) that ensures that radio devices can access the spectrum only according to their privileges (Xu et al., 2006). The framework is based on a *trusted computing* (TC) base or module in each CR node that enforces the policy rules for spectrum access and etiquettes defined in the *XG Policy*

*Language* (XGPL). TRIESSTE has two levels of etiquette enforcement mechanisms. The first enforcement is through an on-board mechanism that ensures trustworthy radio operation by restricting any operation that attempts to violate the policies with the help of a component located in each CR node. In the second level, an external infrastructure consisting of spectrum sensors monitors the radio environment and reports the measurements to the spectrum policy agents. If any violation is detected at any CR node, an appropriate punishment policy is enforced on the offending node.

A robust technique to prevent unauthorized spectrum access in CR network is presented in (NIAP, 2009). The proposed scheme is based on a reliable estimation of the level of the interferences created by the secondary users.

Unauthorized spectrum sensing in a CR network can be prevented by deploying a *spectrum monitoring system* in the network. A typical spectrum monitoring system monitors the spectrum usage in a specific spatial region and over a range of frequencies, and identifies the wireless services and the nodes providing such services. However, design of an effective spectrum monitoring system is challenging since natural or man-made obstacles can change the features of the radio signal and identification of wireless services may be difficult if an attacker can successfully emulate a specific wireless service being provided in the network. To address these issues, spectrum monitoring systems are usually designed and distributed across a number of nodes in a CR network. Information on the wireless services in an area can be transmitted to a *central monitoring location*, which can, then, correlate the various inputs and check the received information against other data like the known position of the wireless services in the area and their source. The major drawback of this approach, however, is that the spectrum sensing capabilities and the amount of data which can be transmitted by the users' devices could be of limited value because of various constraints in the nodes.

Atia et al. propose a model to define an enforcement structure for defending against malicious attacks (Atia et al., 2008) on a CR network. The ultimate goal of the work is to provide an efficient framework so that the primary user is able to distinguish between the uncertain background of wireless environmental losses and the presence of harmful and interference secondary users. In order to minimize interference from the secondary users, the authors propose the use of *silence slots* during which no secondary transmission is allowed and compliance to this norm is enforced at the device certification level. For identifying a device, the authors propose an approach in which the identity of a device is implicitly announced by the *pattern of use/interference* itself. While developing the enforcement framework, the authors note the following fundamental tradeoffs: (i) with the increasing number of potential users to be supported and increasing level of robustness in the system, the time required for detecting malicious users increases, (ii) if the system efficiency in terms of achievable utilization rates is to be increased, the timeliness will degrade, (iii) if a large number of distinct identities for potential users are to be supported, the cost becomes prohibitively large, and hence a gradual punishment mechanism is to be followed in which innocent bystanders may face false conviction for a short period of time. The authors also provide quantifications of the all these three tradeoffs among the various system parameters.

## 6.6 Defense against attacks on cognitive engine

IEEE 802.22 standard provides a robust authentication and encryption scheme to mitigate attacks against the MAC layer. Various mechanisms have also been proposed by researchers for defending against attacks on the cognitive engines of a CR network.

Perich and McHenry propose a policy-based spectrum access control system for the Defense Advanced research Projects Agency (DARPA) NeXt Generation (XG) communications program for mitigating the harmful interference caused by a malfunctioning device or a malicious user for a cognitive *software defined radio* (SDR) (Perich & McHenry, 2009). The authors propose two protection mechanisms for defending against attacks on the cognitive engine. In the first approach, the authors argue that the likely effect of a threat on a CR network is to disrupt the state machine of the CR network and to bring the CR device to an incorrect (i.e. faulty) state. Formal state-space validation, as done with cryptographic network protocols, can be applied to the state machine to ensure that a *bad state* is never arrived at. In the second approach, the authors propose that the beliefs of the cognitive engine should be constantly re-evaluated and compared to *a priori* knowledge (e.g., local spectrum regulations) or rules (e.g., the relationship between transmit power, propagation, and frequency). The authors also present the details of their experimental framework to illustrate the capability offered to radios for enforcing policies and the capability for managing radios and securing access control to interfaces changing the policies of the radio.

In CR networks, reputation and trust-based schemes may be deployed to identify the CR nodes whose cognitive engines are not working as per the policies and rules of the network. Once these nodes are identified, appropriate security policies may be enforced to ensure that the nodes cannot access any network resources till their reputation values increase and reach a minimum acceptable value.

## 6.7 Security mechanisms for cognitive control channels

The *cognitive pilot channel* (CPC) of a CR network is a particularly vulnerable entity. The CPC is responsible for distributing the cognitive control messages to support the CR functions. The CPC is vulnerable to numerous attacks especially the DoS attack by jamming the control channel, and the saturation attack on the control channel.

A popular protection mechanism against the jamming attack in a specific spectrum band of a CR network is to use frequency hopping. The CPC could use more than one spectrum band and "hop" around the spectrum bands to avoid a possible jamming attack. The trade-off is an increased complexity of the CR network as the CR nodes should be notified about the change in the frequency band of the CPC. If an attacker effectively monitors the CPC, it could "chase" the CPC band for every change and eventually cause continual adaptation and outage of service to the CR network. Another issue is the need to allocate various spectrum bands for CPC, which may not be acceptable by spectrum regulators.

For designing an efficient anti-jamming coding technique,

Yue et al. present two coding schemes for recovering lost packets transmitted through parallel channels (Yue & Wang, 2009; Yue et al., 2007). The two coding schemes are known as *rateless coding* and *piecewise coding*. For piecewise coding, the authors present the optimal and several suboptimal designs methods to build short block codes with small number of parity checks. For cognitive radio applications, the authors consider two types of sub-channel selections – *single uniform* and *general non-uniform*. Under both these sub-channel selection strategies, the throughput and the goodput performance of the secondary nodes in a CR network employing either of the anti-jamming coding technique have been analyzed. The results show that both coding techniques provide reliable transmissions with a sustained high level of throughput. The piecewise coding when used with short codes provides better performance with smaller overhead under low to medium jamming condition. For non-uniform sub-channel selection strategy, the short code is found to enhance the throughput and goodput of the secondary transmission with anti-jamming piecewise coding while the rateless coding is found to provide similar or worse results when compared to the uniform case. Both these coding techniques can be applied for protecting the CPC of a CR network.

Meucci et al. present a lightweight mechanism for achieving security in the PHY layer in a CR network using *orthogonal frequency division multiplexing* (OFDM) (Meucci et al., 2009). In this scheme, the user's data symbols are mapped over the physical sub-carriers using a permutation strategy. The security in the PHY layer is achieved using a random and dynamic sub-carrier permutation which is based on a pre-shared information and also on the dynamic spectrum access (DSA) strategy used. The dynamic sub-carrier permutation is allowed to vary with time, change of geographical location and environmental status, providing a very high level of robustness and security. The proposed scheme is effective against eavesdropping attack even if the eavesdropper adopts a long-term pattern analysis. This mechanism can be adapted for protecting CCCs in a CR network although the computational overhead may be prohibitively high in a large-scale CWSN.

## 6.8 Security using geo-location database

In this approach, the CR network provider maintains a database of the positions and transmission characteristics (e.g., transmit power) of all the primary users in the network. The CR finds its own location information using a GPS and compares the data received from the spectrum sensing functionality with the known position of the primary users. Any anomaly in position information triggers an alert for a possible malicious attack. The database containing the information about the primary users and their location can be downloaded form an authenticated and trusted server in the network.

Geo-locations-based security mechanisms in CR networks are simple and do not need sophisticated CR nodes for their operations. However, these techniques are vulnerable to security attacks on *global navigational satellite systems* (GNSS) such as spoofing, or lack of GNSS availability, especially in urban environments (e.g., urban canyons). Borth et al. have proposed a protection technique that is based on beacons emitted by the primary users wherein a primary user would transmit a beacon to alert any secondary user to not transmit in specific spectrum bands (Borth et al., 2008). The disadvantage of this solution is that primary users should modify their equipment to provide the beacon transmission.

Table 2 presents a summary of various possible attacks on CWSNs and their respective possible defense mechanisms.

**Table 2**. **Various security vulnerabilities in CWSNs and their corresponding defense mechanisms**

| Attack Category | Specific Attack Type | Security Mechanism |
|---|---|---|
| Attacks on Communication Protocols | Replay attack | Use of robust authentication scheme. |
| | DoS attack | Use of frequency hopping in the cognitive control channel, code spreading etc. |
| | Malicious alteration of cognitive messages | Protection techniques based on trust or reputation, identification of masquerading threats through signal analysis, authentication of the CR nodes. |
| | Sybil attack | Use of (i) direct validation techniques of nodes including radio resource test, (ii) random key pre-distribution with identity of each node associated with the assigned keys. |
| | Hidden node problem | Data fusion process of collaborative spectrum sensing. |
| | Saturation of cognitive control channel | Robust system design and use of security scheme for protection of system integrity. |
| | Eavesdropping of cognitive radio messages | Protection of message confidentiality. |
| | Disruption of MAC, network and cognitive engine | Verification of the identities, controlled access to the resources, protection of the system integrity. |
| Masquerading Attacks | Primary user emulation attack | Protection techniques based on trust or reputation, identification of masquerading threats through signal analysis, authentication of the CR nodes. |
| | Masquerading of a secondary CR node | |
| Unauthorized Access to Spectrum | Unauthorized use of spectrum band for selfish use by an attacker | A robust and secure framework to enforce spectrum policies. |
| | Unauthorized use of spectrum band for DoS attack on primary users | A robust and secure framework to enforce spectrum policies. |
| Physical Attacks on Sensor Nodes | Physical compromise of the nodes and extraction of cryptographic credentials from the sensor nodes. | Use of (i) tamper-proof sensor hardware, (ii) trusted computing platforms in the sensors. |
| Internal Failure of Sensor Nodes | Physical failure of the sensor node hardware or Byzantine failure of sensor nodes. | Use of collaborative sensing techniques and secure and distributed data fusion process. |
| Power Exhaustion Attacks on Sensor Nodes | Sleep deprivation, Frequent channel change request to drain energy | Use of robust authentication and distributed collaborative sensing. |
| Attacks on the Objective Functions on Sensor Nodes | Belief-manipulation on sensors nodes so that the overall goal optimization module of the network produces suboptimal results. | Use of techniques to enhance robustness in the sensory inputs, resilient and collaborative learning algorithms in the learning phase of the cognitive engine. |
| Attacks on Administrative Policies of Sensor Nodes | Malicious alteration in administrative policies of sensor nodes. | Protection techniques based on trust or reputation, identification of masquerading threats through signal analysis and authentication of the CR nodes. |
| | Excuse attack, Newbie-picking attack. | Use of robust authentication and trusted hardware in sensor nodes. |
| Attacks on Cryptographic Protocols | Malicious attacks on the security and key management protocols. | Use of highly secure protocols with robust key distribution and management schemes. |
| Attacks on Privacy of Sensor Data | Traffic analysis, eavesdropping on sensitive sensor (i.e. source location information) | Use of (i) anonymity mechanisms, (ii) flooding – probabilistic and phantom, and (iii) onion routing. |

# 7. Emerging Research Directions

Wireless technology is rapidly proliferating into all aspects of computing and communications. There are over 8 billion wireless devices in use today (mostly cell phones and mobile computers), and this number is expected to increase to about 100 billion by the year 2025 (Steenkiste et al., 2009). This phenomenal growth in wireless usage will be driven by new applications that embed computing power into the physical world around us, helping us to make the world safer, smarter and more accessible. Radio technology will be at the very heart of the future computing world – one in which billions of communications, mobile devices and sensor/actuators are connected to the global Internet and serve as the foundation for many exciting new classes of applications. However, the anticipated exponential growth of the wireless devices and applications is contingent on our ability to design radio technologies that continue to work well with increasing deployment density – in particular, radio systems must change, and change rapidly, to cope with 2-3 orders of magnitude increase in density from 10-100 devices/km$^2$ today to 1000-10,000 devices/km2 in 2025. Given the fact that spectrum is a finite resource, this call for disruptive technology innovation in the radio field (Steenkiste et al., 2009). Cognitive radios in general and cognitive wireless sensor networks in particular offer the promise of bring just this disruptive technology innovation that will enable the future wireless world. Although CWSN technology have already emerged from the early stage of laboratory trials and vertical applications supports to become a general-purpose programmable radio, there is still a big gap between having a flexible cognitive radio, effectively a building block, and the large-scale deployment of cognitive sensor networks that dynamically optimize spectrum use. Building and deploying a network of cognitive radios is a complex task. The research community working on cognitive radio networks need to understand a wide range of issues including smart antenna technology, spectrum sensing and measurements, radio signal processing, hardware architectures including (SDR) (Li et al., 2009), medium access control (MAC), network discovery and self-organization, routing, adaptive control of mechanisms, policy definitions and monitoring, and learning mechanisms. This is a very wide range of technologies to harness and apply, and hence understanding and properly controlling the behavior of the resulting system is a challenging research task. Given the complexity and multi-dimensional nature of the cognitive radio research, the following research challenges will be important in the immediate future: (i) designing a framework for spectrum policy alternatives and system models, (ii) designing smarter spectrum sensing algorithms, (iii) defining more efficient CR architecture and software abstractions, (iv) designing more efficient cooperative wireless communication systems, (v) design smarter algorithms for dynamic spectrum access, (vi) defining protocol architectures for cognitive networks, (vii) design of cognitive algorithms for adaptation and resource management, (viii) security and privacy issues, (ix) integration and inter-operability issues among CR networks and the Internet.

Since the core focus of this paper is on the security and privacy issues in CWSNs, we identify some of the major security and privacy challenges which need attention from the research community and need to be addressed for large-scale adoption of these networks for real-world deployments. With the advent of CR networks, programmability extends to the radio and hence it becomes possible to create a wide range of authorized and unauthorized waveforms with a low-cost consumer device. It would then be relatively easy to create denial-of-service attacks that can affect critical applications such as traffic control or healthcare. The regulatory bodies need to be aware of these potential threats and work with industry to develop trusted hardware architectures, monitoring frameworks or other solutions to the security problems. Some of the issues that need attention in this regard are as follows:

- What types of denial-of-service and other security attacks are made possible by emerging cognitive radio technology?
- Software weaknesses are known to be a major security problem in the Internet today – what are the implications of increasingly software-based radio implementations?
- How does one assure that CRs operate as intended and designed? Is there a trusted cognitive radio architecture which can address some of these security concerns?
- What authentication mechanisms are needed to support cooperative cognitive networks? Are

reputation-based schemes useful supplements to conventional PKI authentication protocols?

- How the current protection techniques for spectrum management and spectrum sharing functions can be further improved? What link protection techniques could be further incorporated in the current security frameworks?
- How the performance efficiency of the different protection techniques used in collaborative spectrum sensing can be evaluated in real-world deployment scenario?
- How to design and standardize tamper-resistant module to enforce spectrum regulation policies in CR nodes and SDR devices?

At the same time, cognitive radios offer important new capabilities to defend against intrusions or denial-of-service attacks. The spectrum sensing and SDR capability of the radio make it feasible to employ recent developments in wireless security in which physical layer properties (such as RF signatures) are used for authentication or secure communication (Mathur et al., 2008). Also, spectrum scanning and agility associated with cognitive radios enable networks to move away from frequency channels experiencing denial-of-service attack. Location is another important feature of a wireless network, and information on geographic position can also be used to defend against certain types of attacks on cognitive networks. Some of the research issues which need to be addressed in this regard are as follows:

- Identification of physical layer security enhancements for wireless networks, and evaluation of performance in real-world deployment scenarios.
- Evaluation of denial-of-service attack scenarios and method for defense.
- Use of geo-location for improved wireless network security
- Cooperative methods for detecting and isolating intruders.

While the ongoing research works on these issues are quite promising, evaluations have been mostly limited to lab environments, and it is not clear to what degree these techniques will be feasible in real-world deployments, or whether these algorithms, architectures and protocols will scale to high density environments. Large scale testing in the CR networks is mandatory for this purpose. Since the CWSNs are still in their pre-deployment phase, there is still an opportunity and a critical requirement to make security as an integral component of CR network architecture. This will require realistic practical evaluation of new techniques as they are designed and developed.

## 8. Conclusion

The cognitive radio paradigm introduces entirely new types of security threats to wireless networks in general and wireless sensor networks in particular. It makes the development of effective security models and mechanism very challenging. However, wireless security in cognitive radio networks is a technical area that has received relatively less attention, even though security will likely to play a key role in the long-term commercial viability of the technology. This paper has first introduced various security threats in traditional wireless sensor networks which are also applicable in cognitive wireless sensor networks. It has then identified various additional security threats which are applicable for cognitive wireless sensor networks and presented various challenges in defending against these security vulnerabilities. In addition to identifying various threats, the paper has also discussed various existing security mechanisms to defend against these threats and attacks. A comprehensive taxonomy of the attacks and their respective security schemes are also presented. Some key research challenges in CR networks particularly from the perspectives of security and privacy are identified and discussed briefly. These challenges need to be addressed in the near future by the research community in order to make deployment of CWSNs feasible in critical and sensitive real-world applications.

## Acknowledgement

The author gratefully acknowledges the contributions in this work received from his students Swati Choudhury, Swetashree Mishra, Isha Bharati, Ramesh Kumar, Aiswaraya Mohapatra and Prateek Kumar Nayak of National Institute of Science and Technology (NIST), Odisha, India.